\documentclass[twocolumn]{aastex631}

\renewcommand{\textsc}[1]{#1}

\usepackage{float}
\usepackage{amsmath}
\usepackage{siunitx}
\usepackage{xspace}
\usepackage{blindtext}
\usepackage{paracol}
\usepackage{booktabs}
\usepackage[utf8]{inputenc}

\newcommand{\com}[1]{}

\newcommand{\zw}{II\,ZW\,096\xspace}
\newcommand{\zwA}{II\,ZW\,096A\xspace}
\newcommand{\zwB}{II\,ZW\,096B\xspace}

\begin{document}
%\tablenum{42}
%\linenumbers % Activar números de línea

\title{High-Resolution Optical IFU Spectroscopy of the Complex Galaxy Merger II Zw 096}

\author[0009-0005-7950-3288]{Clemente Riesco}
\affiliation{Instituto de Astrof\'isica, Facultad de F\'isica, Pontificia Universidad Cat\'olica de Chile, Casilla 306, Santiago 22, Chile}

\author[0000-0001-7568-6412]{Ezequiel Treister}
\affiliation{Instituto de Alta Investigaci{\'{o}}n, Universidad de Tarapac{\'{a}}, Casilla 7D, Arica, Chile}

\author[0000-0001-8349-3055]{Giacomo Venturi}
\affiliation{Scuola Normale Superiore, Piazza dei Cavalieri 7, 56126, Pisa, Italy}

\author[0000-0002-8686-8737]{Franz E. Bauer}
\affiliation{Instituto de Alta Investigaci{\'{o}}n, Universidad de Tarapac{\'{a}}, Casilla 7D, Arica, Chile}

\author[0000-0003-3474-1125]{George C. Privon}
\affiliation{National Radio Astronomy Observatory, 520 Edgemont Road, Charlottesville, VA 22903, USA}
\affiliation{Department of Astronomy, University of Virginia, 530 McCormick Road, Charlottesville, VA 22903, USA}
\affiliation{Department of Astronomy, University of Florida, P.O. Box 112055, Gainesville, FL 32611, USA}

\author[0000-0003-1778-1061]{Carolina Finlez}
\affiliation{Instituto de Astrof\'isica, Facultad de F\'isica, Pontificia Universidad Cat\'olica de Chile, Casilla 306, Santiago 22, Chile}

\author[0000-0003-4546-897X]{Sandra Zamora}
\affiliation{Scuola Normale Superiore, Piazza dei Cavalieri 7, 56126, Pisa, Italy}

\author[0000-0002-2688-7960]{Dusan Tub\'in-Arenas}
\affiliation{Leibniz-Institut f\"ur Astrophysik Potsdam (AIP), An der Sternwarte 16, 14482 Potsdam, Germany}
\affiliation{Potsdam University, Institute for Physics and Astronomy, Karl-Liebknecht-Straße 24/25, 14476 Potsdam, Germany}

\author[0000-0002-3139-3041]{Yiqing Song}
\affiliation{European Southern Observatory, Alonso de C\'ordova, 3107, Vitacura, Santiago 763-0355, Chile}
\affiliation{Joint ALMA Observatory, Alonso de C\'ordova, 3107, Vitacura, Santiago 763-0355, Chile
}
\author[0000-0001-8931-1152]{Ignacio del Moral-Castro }
\affiliation{Instituto de Astrof\'isica, Facultad de F\'isica, Pontificia Universidad Cat\'olica de Chile, Casilla 306, Santiago 22, Chile}

\author[0000-0001-5231-2645]{Claudio Ricci}
\affiliation{Instituto de Estudios Astrof\'isicos, Facultad de Ingenier\'ia y Ciencias, Universidad Diego Portales, Av. Ej\'ercito Libertador 441, Santiago, Chile}
\affiliation{Kavli Institute for Astronomy and Astrophysics, Peking University, Beijing 100871, People’s Republic of China}

\author[0000-0001-8353-649X]{Cristina Ramos Almeida}
\affiliation{Instituto de Astrof\'isica de Canarias, Calle V\'ia Lactea, s/n, E-38205 La Laguna, Tenerife, Spain}
\affiliation{Departamento de Astrof\'isica, Universidad de La Laguna, E-38206 La Laguna, Tenerife, Spain}

\author[0000-0003-4209-639X]{Nancy A. Levenson}
\affiliation{Space Telescope Science Institute, 3700 San Martin Drive, Baltimore, MD
21218, USA}

\author[0000-0002-1912-0024]{Vivian U}
\affiliation{4129 Frederick Reines Hall, Department of Physics and Astronomy, University of California, Irvine, CA 92697, USA
}

\author[0000-0001-7421-2944]{Anne M. Medling}
\affiliation{Department of Physics \& Astronomy and Ritter Astrophysical Research Center, University of Toledo, Toledo, OH 43606, USA}

\author[0000-0002-5828-7660]{Susanne Aalto}
\affiliation{Department of Space, Earth and Environment, Chalmers University of Technology, SE-41296 Gothenburg, Sweden}

\author[0000-0001-9697-7331]{Giuseppe D'Ago}
\affiliation{Institute of Astronomy, University of Cambridge, Madingley Road,
Cambridge CB3 0HA, UK}

\author[0000-0001-6638-4324]{Valeria Olivares}
\affiliation{Departamento de F\'isica, Universidad de Santiago de Chile, Av. Victor Jara 3659,
Santiago, Chile.}

\author[0000-0003-0057-8892]{Loreto Barcos Muñoz}
\affiliation{National Radio Astronomy Observatory, 520 Edgemont Road, Charlottesville, VA 22903, USA}
\affiliation{Department of Astronomy, University of Virginia, 530 McCormick Road, Charlottesville, VA 22903, USA}

\author[0000-0001-5742-5980]{Federica Ricci}
\affiliation{Dipartimento di Matematica e Fisica, Universita Roma Tre, via della Vasca Navale 84, I-00146 Roma, Italy}

\author[0000-0003-4830-9069]{Gustav Olander}
\affiliation{Department of Space, Earth and Environment, Chalmers University of Technology, SE-41296 Gothenburg, Sweden}

\author[0000-0002-2713-0628]{Francisco Muller-Sanchez}
\affiliation{Department of Physics and Materials Science, The University of Memphis, 3720 Alumni Avenue, Memphis, TN 38152, USA}

\author[0000-0001-5242-2844]{Patricia B. Tissera}
\affiliation{Instituto de Astrof\'isica, Facultad de F\'isica, Pontificia Universidad Cat\'olica de Chile, Casilla 306, Santiago 22, Chile}
\affiliation{Centro de Astro-Ingenier\'ia, Universidad Cat\'olica de Chile, Casilla 306, Santiago 22, Chile }
%\author[0000-0001-6920-662X]{Neil Nagar}
%\affiliation{Departamento de Astronom\'ia, Universidad de Concepci\'on, Concepci\'on, Chile}

%% automatically takes care of all commas and "and"s between authors' names.


\begin{abstract}
Luminous and Ultra-luminous IR galaxies ((U)LIRGs) are critical for investigating feedback mechanisms due to a combination of intense star formation (SF) episodes and active galactic nuclei (AGN), particularly in the context of complex galaxy interactions. We conduct a detailed analysis of the  \zw merging system using the Multi-Unit Spectroscopic Explorer (MUSE) on the Very Large Telescope (VLT), combining high-resolution Narrow Field Mode (NFM) and large-area Wide Field Mode (WFM) observations. We mapped the morphology, kinematics, and ionizing radiation of the system's gas by fitting atomic emission lines and the optical continuum. We identify three or more distinct galaxies within \zw, revealing rotational patterns and complex interactions consistent with a collapsing small galaxy group. The kinematics and ionization structures suggest high star formation rates and shock-driven processes, which align with this proposed scenario. Focusing on the D1 compact region, which contributes 40–70\% of the system's IR emission, and combining information from archival multi-wavelength observations, we find strong evidence of a heavily obscured AGN powering it. Our analysis of the internal structure, interactions, and merger state of \zw offers novel insights into the galaxy evolution processes in this dynamic and highly chaotic system.

\end{abstract}

%% You will be asked to select these concepts during the submission process
\keywords{infrared: galaxies -- galaxies: interactions -- galaxies: individual (\zw) -- techniques: spectroscopic}

%% and \citet commands to identify citations. The citations are

\section{Introduction} \label{sec:intro}

Major mergers play a crucial role in the context of hierarchical galaxy evolution. These interactions produce features such as bridges and tidal tails due to gravitational forces (see e.g. \citet{Canalizo2007, Bennert2008, RamosAlmeida2011, Pierce2023}), as predicted by merger simulations \citep{Toomre1972,Quinn1984, Hernquist1992, Cattaneo2005, Lotz2008, Feldmann-Carollo-Mayer2010}. Inside galaxies, gravitational torques cause gas inflow to the center \citep{Mihos1995}, generating starbursts (SB) and triggering the active galactic nuclei (AGNs). Hence, the kinematics of galaxy mergers are critical to understanding the physical processes in this phase, which can significantly change a galaxy's evolution. 

The interactions between galaxies are critical in the 'cosmic cycle' model \citep{Sanders1988}, as they can trigger bursts of star formation, induce feedback processes that alter fundamental properties such as metallicity and gas content, and drive morphological transformations through mergers and tidal interactions \citep{DiMatteo2005, Kewley2013}. \cite{Hopkins2005} describe the different stages transited by a galaxy undergoing a major merger. It begins with merging dark matter (DM) halos containing isolated galaxies. Once in a single merged halo, galaxies begin to interact, losing angular momentum and temporarily increasing their Star Formation Rate (SFR). As the distance between nuclei decreases, galaxy disks collide, igniting intense star formation (SF) and AGN activity due to increased gas inflow to the center. The large amounts of centrally concentrated dust absorb the optical and UV emission from young stars and the AGN, re-emitting it at IR wavelengths, leading to the Luminous or Ultra Luminous Infrared Galaxy ((U)LIRG) phase (LIRG; $L_{IR}>10^{11}L_{\odot}$ and ULIRG; $L_{IR}>10^{12}L_{\odot}$). This merged galaxy could undergo a luminous quasar phase, initially hosting an obscured AGN, which, once the supermassive black hole (SMBH) has grown sufficiently, would expel the material via feedback action \citep{Granato2004, Brandt2005, Granato2006, Lapi2006, HPR2007}, until the SFR decreases due to the lack of gas or to the negative feedback caused by energetic sources (AGNs, stars and supernovae). After reaching a semi-virialized state, the resulting galaxy would be classified as a massive elliptical galaxy with low or non-SF activity. 
In this scenario, (U)LIRGs represent a critical step. They are among the most luminous and active galaxies in the Local Universe, with SFR reaching up to $\sim$100 $M_\odot\ / \rm yr$, meaning that they are rapidly growing their stellar bulge and the SMBH \citep{Armus2009}. Their active and fast nature makes them perfect candidates for understanding the interplay between the increased SF and SMBH growth. 

The Great Observatories All-sky LIRG Survey \citep[][GOALS]{Armus2009} is a multiwavelength survey covering radio to X-ray observations, incorporating data from various facilities to study over 200 local (U)LIRGs pre-selected from the IRAS Revised Bright Galaxy Sample (RBGS; \citealp{Sanders2003}), spanning a wide range of optically classified galaxies: AGN, low-ionization nuclear emission-line region(LINER), SB; and interaction stages: minor mergers, major mergers, and non-interacting galaxies. As such, GOALS provides a diverse sample of galaxies with enhanced infrared emission in the local universe, enabling the study of physical processes with higher resolution. 

\zw is one of the LIRGs from the GOALS sample. It is a local ($z=0.0365$) merging system with $L_{\text{IR}} = 8.7 \times 10^{11}\,L_{\odot}$ \citep{Armus2009}. According to the merger classification system of \citet{Stierwalt2013}, this system is in the c stage: strong tidal tails, amorphous disks, and other signs of merger activity preceding the coalescence of both nuclei into a single envelope. This merger was previously identified with four IR peaks \citep{Goldader1997}: two of these, \zwA and \zwB, have been identified as galaxies based on their morphology, while the other two regions, designated C and D, remain unclassified, as can be seen in Figure \ref{fig:Fig1}. Consequently, the system is classified as a merger of two galaxies.

Previous Spitzer observations revealed that up to 80\% of the L$_{8-1000 \mu m}$ comes from the D region, outside of the central galaxies(\zwA and \zwB) in the merger \citep{Inami2010}. Higher resolution James Webb Space Telescope (JWST) observations \citep{Inami2022} aid our understanding of the key physical regions, indicating that 40–70\% of the 8–1000$\mu$m emission originates from a compact region designated as D1, which has a radius of no more than 175 pc at this wavelength and less than 70 pc at 33 GHz \citep{BarcosMunoz2017}. They used the 6.2 $\mu$m PAH EQW diagnostic, distinguishing starbursts—characterized by high EQW due to strong PAH emission—from AGNs, where a dominant hot dust continuum and PAH destruction yield low EQW. The intermediate EQW measured in D1 indicates that this region may harbor a dust-obscured AGN.

\citet{Ricci2021} analyzed this system in X-rays using Chandra and XMM-Newton, finding that the entire system, including both galaxies and the additional IR peaks, was well-fitted by a star formation model. Consequently, this analysis cannot be used to classify D1, as the contribution from the other merger sources contaminates its classification. Additionally, their study reported no detection of D1 with NuSTAR, meaning that the presence of a heavily obscured AGN cannot be excluded. However, this non-detection sets an upper limit on the AGN luminosity, which depends on the unknown column density. Hence, the dynamics and composition of this system, especially in D1, are still not fully understood.

In this work, we present VLT/MUSE data for the merging system \zw to study the ionized gas's spatial distribution and physical properties, including its excitation mechanism and kinematics. By combining this information with Chandra X-ray and ALMA Band 3 data, we aim to understand the nature of regions like D1 and reconstruct the system’s dynamical stage, providing insight into its structure, interactions, and the number of galaxies involved. This work is organized as follows: Section \ref{sec:Data description and Analysis} details the IFU data and analysis. Section \ref{sec:morphological description} describes the system’s morphology. In Section \ref{sec:Emission line analysis}, we present the emission line flux, velocity, and velocity dispersion maps along with ionization diagnostic diagrams. Section \ref{sec:Chandra analysis} examines Chandra X-ray data for ionization classification. Finally, Sections \ref{sec:Discussion} and \ref{sec:Conclusion} contain the discussion and conclusions. In this paper, we assume a $\lambda$CDM cosmology with $h_{0}=0.7$, $\Omega_{m}=0.27$ and $\Omega_{\lambda}=0.73$ \citep{Hinshaw2009}.
%Besides that, in this paper the east region is referred to as the one containing the regions C and D, the south galaxy to the A region, and the north-west (NW) galaxy to the B region, all this based on the FoV of the NFM observations.

\section{Data description and Analysis} 
\label{sec:Data description and Analysis}

\begin{figure*}[!ht]
    \centering
    \includegraphics[width=1\linewidth]{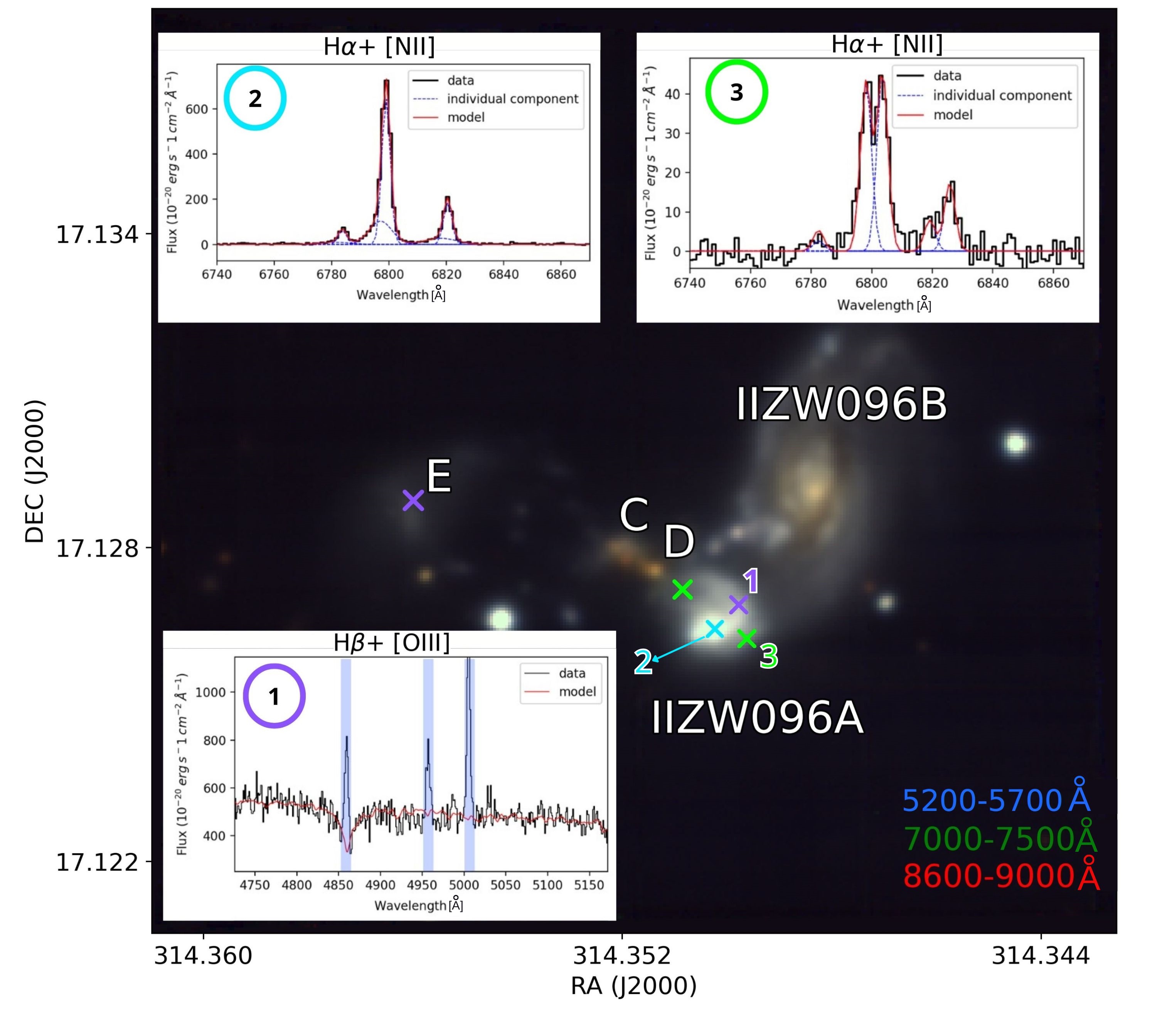}
    \caption{Color composite image of the MUSE WFM data with examples of spectral fits from selected regions. The blue, green, and red channels correspond to the continuum collapse in the [5200-5700]\AA, [7000-7500]\AA, and [8600-9100]~\AA\ ranges, respectively. Spectrum 1 is a Balmer absorption spectrum from the H$\beta$ + [OIII] of the tidal tail of \zwA in the direction of C+D, also representative of the E region. The black lines represent the spectra, the red lines show the spectral fit, and the blue regions highlight the masked emission lines. Spectra 2 and 3 show fits of continuum-subtracted emission lines, requiring two Gaussian components in the H$\alpha$ + [NII] region. The black lines show the spectra, the blue dashed lines represent the individual Gaussian fits, and the red lines show the total model, combining the contributions from both Gaussians. Spectrum 2 corresponds to the emission from the center of \zwA, extracted from the NFM-South, marked by the cyan cross, using both narrow and broad Gaussians for the fit. Spectrum 3 uses two narrow Gaussians (with similar $\sigma$), represented by the green crosses. It corresponds to the region where \zwA overlaps with \zwB, also representing the overlap between \zwA and the C+D region.}
    \label{fig:Fig2}
\end{figure*}

\begin{figure*}[!htb]
    \centering
    \includegraphics[width=1\linewidth]{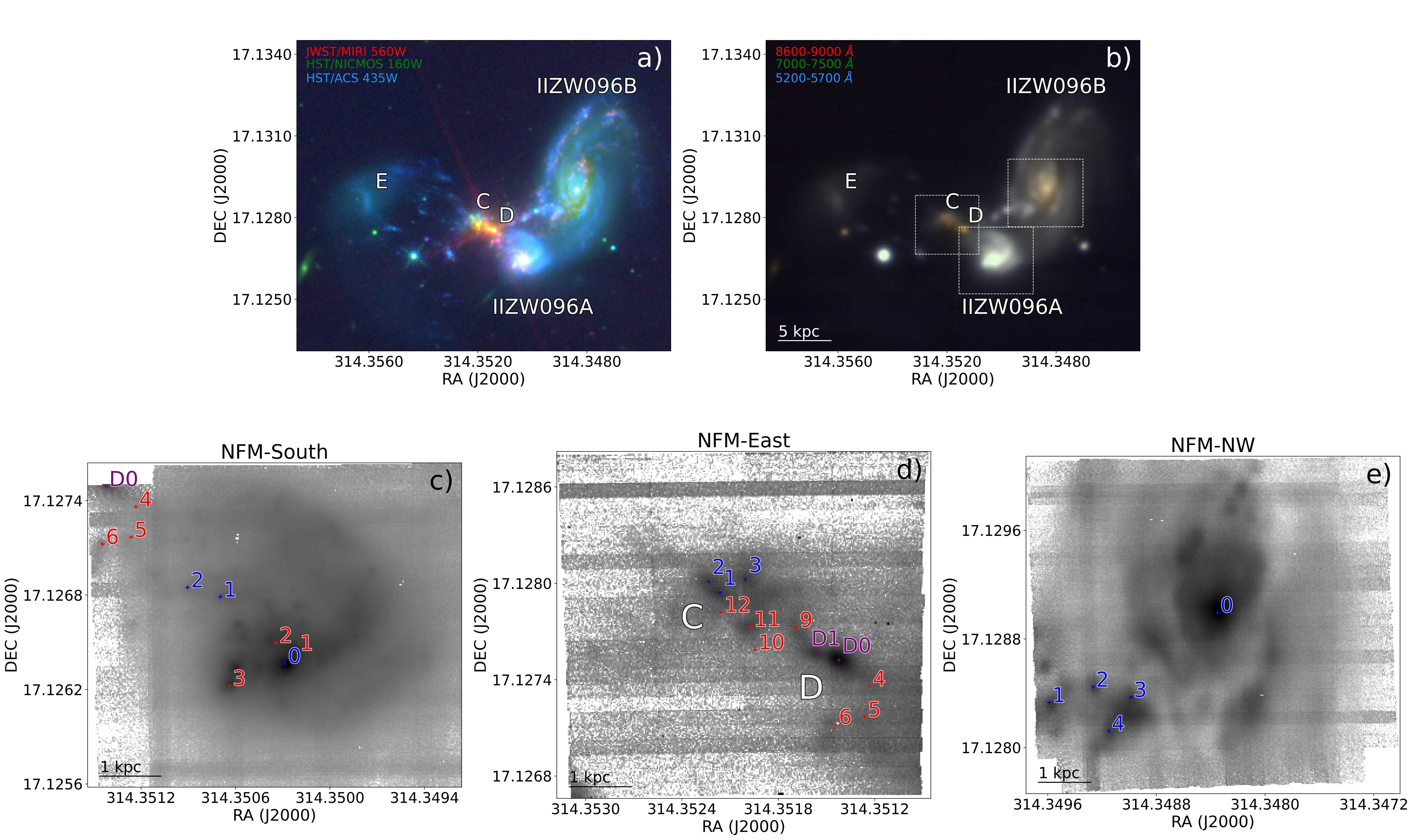}
    \caption{Multiple observations of \zw. In all panels, the region's names follow the nomenclature of \citet{Goldader1997}. (a) Composite color image of \zw\ using JWST/MIRI F560W (red), HST/NICMOS F160W (green), and HST/ACS F435W (blue). (b) Color image from Figure \ref{fig:Fig2}, where blue, green, and red correspond to images collapsed over the following emission-line-free wavelength ranges: [5200-5700]~\AA, [7000-7500]~\AA, and [8600-9100]~\AA . The dashed squares indicate the FOV of the three NFMs: the one centered on \zwA\ is the NFM-South, the one containing the C+D region is the NFM-East, and the one centered on \zwB\ is the NFM-NW. (c)–(e) MUSE/NFM white-light images covering the [8600-9100]~\AA\ range. The red sources correspond to JWST/MIRI observations from \citet{Inami2022}, labeled with an ID not shown on the map to reduce visual clutter, while the blue sources are CO/ALMA detections from \citet{Wu2022}. In the latter, source names are preceded by the letters A, B, and C, corresponding to \zwA in NFM-South, \zwB in NFM-NW, and the C+D region in NFM-C, respectively. The sources detected in both emissions are D0 and D1, which are referred to by their full names due to their relevance in this work. }
    \label{fig:Fig1}
\end{figure*}

\subsection{Data description and basic reduction}
\zw was observed with MUSE in the Wide Field Mode (WFM) on September 23, 2016 (program ID 097.B-0427; PI G. Privon) and later in the Narrow Field Mode (NFM) assisted by Adaptive Optics (AO) on July 2, August 24, 27, and 30, 2022 (program ID 109.23AX.001; PI E. Treister). The WFM seeing-limited observations consist of four exposures with integration times of 800s each, for a total of 3200s on-source. The NFM observations consist of three different pointings, dubbed NFM-South, NFM-East, and NFM-NW, centered on the brightest sources in the system. All the NFM observations had exposure times of 600s each. NFM-NW comprises six exposures, NFM-South seven exposures, and NFM-East one, for a total exposure time of 3600 s, 4200 s, and 600 s, respectively. For the NFM-East pointing, two exposures were available (taken on 2022-08-25T04:24:41.038 and 2022-08-28T03:59:19.123), but we employed only the former and discarded the latter due to its much inferior data quality in terms of spatial resolution and thus signal-to-noise ratio. 
For both the WFM and NFM observations, subsequent exposures were rotated by 90$^o$ and slight offsets in RA and Dec to improve the data reduction process. The spatial offsets were $\pm$0.5\arcsec\  for the WFM data and $\pm$0.1\arcsec\ for the NFM observations.

We performed the data reduction using the ESO MUSE pipeline (\citealt{Weilbacher:2020a}; versions 2.6 and 2.8.1 for the WFM and NFM data, respectively) by employing the software ESO Reflex (Recipe flexible execution workbench; \citealt{Freudling:2013a}), that gives a graphical and automated way to execute the Common Pipeline Library (CPL; \citealt{Banse:2004a, ESOCPL2014}) reduction recipes with EsoRex (ESO Recipe Execution Tool; \citealt{ESOCPL2015}), within the Kepler workflow engine \citep{Altintas:2004a}. The pipeline reduction corrects the data for bias, flat field, illumination, wavelength calibration, flux calibration, geometric reconstruction of the data cube, sky subtraction, and exposure combination.
In the case of the WFM data, the sky background to be subtracted was determined from within the scientific field of view (FOV) regions free of emission from the target.
In the case of the NFM data, given the smaller FOV, dedicated offset observations of the sky background, $\sim$30-40$''$ away from the target, were obtained and used for the sky subtraction.
The FWHM seeing of the combined WFM cube varied across the wavelength range, with values between 0.76-0.94'', which correspond to 0.55-0.74 kpc at the distance of the system. For the NFM-AO observations, no point-like source was detected in the field of view. Hence, a representative spatial resolution of $\sim$0.1$''$, corresponding to $\sim$80 pc was assumed. The measured galactic E(B–V) value is 0.083 \citep{Schlegel1998}. Given the minimal level of extinction implied, this correction would be negligible and hence not applied.

\subsection{Data analysis}
\textsc{Figure~\ref{fig:Fig2} presents a color composite of the MUSE WFM data. The blue, green, and red channels correspond to the continuum collapsed in the [5200–5700]~\AA, [7000–7500]~\AA, and [8600–9100]~\AA\ wavelength ranges, respectively. The purple cross indicates the E region and the tidal tail of \zwA. An example of a characteristic spectrum from this region is shown, displaying clear Balmer absorption features from the stellar population. This demonstrates that, prior to fitting the emission lines, it was necessary to account for this absorption. These corrections are essential to accurately determine the emission line fluxes, as neglecting the Balmer absorption would lead to a significant underestimation of the flux for certain lines}. For this purpose, the penalized pixel-fitting (pPXF) package \citep{Cappellari2016} was employed. The pPXF package fits the spectra using a linear combination of single stellar population (SSP) templates, incorporating multiplicative and additive Legendre polynomials to account for mismatches between the templates and the observations. Contaminated regions in the spectra, such as emission lines and areas affected by adaptive optics (AO) in the NFM, were masked during the fitting process. The extended MILES (E-MILES) \citep{Vazdekis2016} SSP templates were used due to their large spectral coverage [1680-50000]~\AA. The pPXF package was executed within the GIST (Galaxy IFU Spectroscopy Tool) pipeline \citep{Bittner2019}.

The WFM and NFM data cubes were astrometrically aligned using Gaia Data Release 3 \citep{Brown2021}. Due to the large FoV of the WFM, multiple sources were available for calibration. In the NFM, we used the sources D0, B2, and A0 (identified as sources by Gaia) for the East, NW, and South paintings, respectively. \textsc{The achieved astrometric accuracy in RA and DEC after the alignment is the following: for the WFM 0.01\arcsec; for the NFM South 0.013\arcsec; for the NFM NW 0.091\arcsec; and for the NFM East 0.013\arcsec.}

To obtain a more accurate estimation of the emission line flux, we follow the procedure described by \cite{Venturi2018}. Initially, a Voronoi tessellation \citep{Cappellari2009} is applied to a region free of emission lines in the wavelength range of 7000-7300 \AA, aiming for a minimum S/N of 30 for spaxels with S/N $>$1. This method combines adjacent spaxels until the desired S/N threshold is reached. pPXF is applied to the binned cube from the tesselation, generating a model of the continuum of the spectra for each bin, which is subtracted spaxel-by-spaxel from the original data cube by rescaling it to the observed continuum flux at the spaxel level to mitigate any bin-to-spaxel flux discrepancy, resulting in a continuum-free, non-binned cube.

A new Voronoi tessellation is then applied to this cube, using a given emission line as a reference. Depending on the specific goal, different lines are used to obtain the required S/N. For the optical diagnostic diagrams \citep{Baldwin1981, Veilleux1987}, the H$\beta$ line was used for the tessellation, ensuring a minimum S/N of 3 while excluding spaxels with S/N $\ < $ 1. The instrument has lower spectral resolution and efficiency at bluer wavelengths, so the tessellation was performed on H$\beta$, the shortest wavelength line covered by MUSE, to generate this diagram, ensuring more reliable results for this diagnostic. A second tessellation was performed using H$\alpha$, with a minimum S/N of 3, while excluding spaxels with S/N$\ < $1. This line is typically the brightest in the spectra, resulting in smaller bins and thus reducing the decrement in spatial resolution due to the tessellation. This binning is used to analyze the system's morphology, kinematics, and SFR. Although the [N II] $\lambda$6549, $\lambda$6583 doublet often poses challenges to measure accurately in optical spectra due to its proximity with H$\alpha$, in this case, the line velocity dispersions are sufficiently narrow to successfully separate them without degeneracies in the fit.

The fitted emission lines are then: H$\beta$ $\lambda$4861, [O III] $\lambda$4959, $\lambda$5007, [O I] $\lambda$6300, H$\alpha$ $\lambda$6563, [N II] $\lambda$6549, $\lambda$6583, and [S II] $\lambda$6717, $\lambda$6730. The Python Spectroscopic Toolkit package \citep{Ginsburg2022} was used to perform the emission line fitting. To reduce the degrees of freedom in the fits, the [O III] and [N II] doublets ratios were fixed to their theoretical values of 1$:$3 \citep{Osterbrock1989}.

Complex line profiles were observed in the system, requiring more than one Gaussian component to fit them. \textsc{ In Figure \ref{fig:Fig2}, we present different cases for the Gaussian fits}. For the WFM and NFM-South, three cases were considered for fitting the lines: (1) a single Gaussian component, (2) two Gaussian components with similar $\sigma$ but distinct velocities, and (3) two components, referred to as narrow and broad, where the latter has a higher $\sigma$ and lower amplitude. The second case is presented by the region where \zwA appears to overlap with the C+D regions and \zwB, where the superposition of structures generates these line profiles. The cyan cross near the center of \zwA marks the third case. In the NFM-East, two scenarios were identified: (1) a single Gaussian component and (2) two Gaussian components with similar $\sigma$ but distinct velocities. The second scenario is the same as previously presented by the green cross, where the C+D region and \zwA interact, possibly due to the superposition of structures. The NFM-NW is well-modeled with a single Gaussian component.

The reduced $\chi^{2}$ criterion was used to discriminate between the best Gaussian combinations. A two-Gaussian component model was selected if its reduced $\chi^{2}$ was smaller than 80\% of the value resulting from the one-Gaussian component model. In the case of two possible two-Gaussian combinations, such as for the NFM-South, the model with the smaller reduced $\chi^{2}$ was chosen. Further details on the kinematics are provided in Section \ref{sec:Kinematics}.

%\subsection{\gv{TO BE MOVED TO DEDICATED (SUB-)SECTION}}
\section{Morphological Description of the System}
\label{sec:morphological description}

\zw is a complex system whose interactions are not yet fully understood. To gain better insight into this merger, Figure \ref{fig:Fig1} presents a series of images of the system. Figure \ref{fig:Fig1}a shows an RGB image composed of HST/ACS F435W (optical; blue) (PID 11196, PI A. Evans, \citep{Inami2010}), HST/NICMOS F160W (NIR; green) (PID 11196, PI A. Evans, \citep{Inami2010}), and JWST/MIRI F560W (MIR; red)(PID 1328, PI Armus, L., \citep{Inami2022}). The four IR peaks identified by \citet{Goldader1997}—A, B, C, and D—are shown, along with the low-surface-brightness region outside the main interaction, E. As A and B have been identified as galaxies, we refer to them as \zwA and \zwB.

\zwA is prominent for its bright optical (blue) emission and contains multiple bright clumps in the MIR near its center, possibly due to UV radiation from highly star-forming clumps that is absorbed and re-emitted at this wavelength. The color of \zwB is dominated by optical (blue) emission, with an increasing MIR contribution (red) in the clumps along its spiral arms, similar to \zwA. At its center, enhanced NIR emission (green) is observed, possibly associated with an older stellar population or an increased dust content.

To the east, a prominent tidal feature links \zwA with the C+D region, where the majority of the MIR luminosity (red) is emitted, hosting multiple bright sources identified in both the NIR and MIR. Approximately 7 kpc from the C+D region lies the E region, characterized by significantly lower brightness \citep{Goldader1997}. Between E and C+D, multiple optical sources are detected, suggesting a possible connection between these two regions as remnants from an interaction.

Figure \ref{fig:Fig1}b presents the same color composite image from Figure \ref{fig:Fig2}. \zwA is the brightest galaxy, with no prominent emission in red, consistent with high SF with lower obscuration. The C+D region and the center of \zwB are dominated by red, which may result from an increased dust content or dust-obscured star formation. The dashed white rectangles indicate the FoV of the three NFM cubes: NFM-South, NFM-East, and NFM-NW.

Figures \ref{fig:Fig1}c, \ref{fig:Fig1}d, and \ref{fig:Fig1}e are continuum images collapsed over the same wavelength range as the red color in Figure \ref{fig:Fig1}b. 
\textsc{In these images, the red sources denote compact detections by JWST/MIRI F560W \citep{Inami2022}, at a spatial resolution of 0.18\arcsec. They are labeled with an ID, but these labels are not shown on the maps to reduce visual clutter. The blue sources correspond to CO/ALMA detections \citep{Wu2022}, with a spatial resolution of 0.16\arcsec, while the purple sources are detected in both datasets.} Figure \ref{fig:Fig1}c shows the optical continuum image of \zwA.It hosts several clumps, with the brightest ones detected by IR emission.
Figure \ref{fig:Fig1}d presents the continuum emission of the C+D region. The brightest sources are D0 and D1, the IR bump. To the northeast, the C region contains multiple sources detected in CO and IR.
Figure \ref{fig:Fig1}e shows the optical image of \zwB. Some compact sources are detected in CO in the direction of the C+D region.

\section{Emission line analysis}
\label{sec:Emission line analysis}

\subsection{Morphology}
\label{sec:Morphology}

\begin{figure*}[!ht]
    \centering
    \includegraphics[width=1\linewidth]{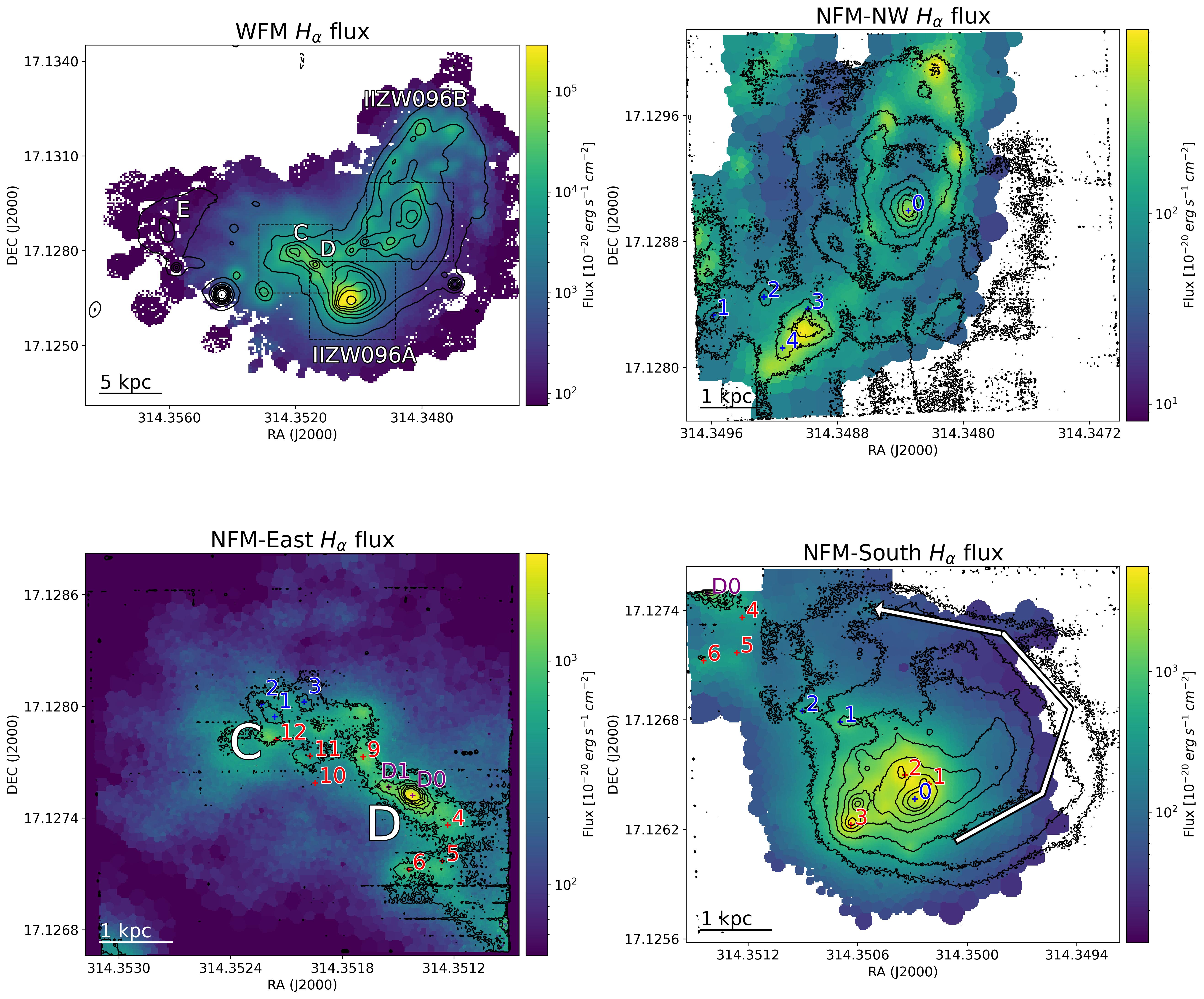}
    \caption{\textsc{Maps of the H$_{\alpha}$ line flux per spaxel for the II ZW 096 system.} Contours represent the collapsed emission in the [8600-9100]\AA\ range, with scale bars of 5 kpcs in the WFM and one kpc in the NFM. The five regions are labeled by their names \zwA, \zwB, C, D, and E. In the NFM, the red sources correspond to JWST/MIRI observations from \citet{Inami2022}, while the blue sources are CO/ALMA detections from \citet{Wu2022}. In the latter, source names are preceded by the letters A, B, and C, corresponding to \zwA in NFM-South, \zwB in NFM-NW, and the C+D region in NFM-East, respectively. The sources detected in both JWST/MIRI and CO/ALMA emissions are D0 and D1, which are referred to by their full names due to their relevance in this work. \textsl{Top left:} Map of the WFM cube, with the NFM FoV represented by dashed rectangles. \textsl{Top right:} Map of NFM-South, covering \zwA, where new sources near the nucleus and in the interacting region are resolved. The white arrow represents the tidal tail extension in the C+D direction. \textsl{Bottom left:} Map of NFM-East, showing the resolved sources from C and D. \textsl{Bottom right:} Map of NFM-NW, revealing multiple sources near the center and in the interaction regions.}
    \label{fig:mom0_halpha}
\end{figure*}

The morphology and kinematics of the system are analyzed using the flux, velocity, and velocity dispersion maps from the total Gaussian fits (Figure \ref{fig:Fig2} ). The Voronoi tessellation based on H$\alpha$ was used to balance spatial resolution and sensitivity within the cube, ensuring a minimum S/N of 3 while excluding spaxels with S/N$<$1. The flux map allows for a detailed analysis of the system's morphology. Figure \ref{fig:mom0_halpha} shows \textsc{maps of measured H$\alpha$ emission per spaxel} from both the WFM and NFM data, where the contours represent the continuum emission in the [8600-9100]~\AA range. Several tidal tails and disrupted regions are observed, typical features of the gravitational interactions in this system, revealing its complexity.

In the WFM, the brightest source is \zwA, which, as shown in Figure \ref{fig:Fig1}, is dominated by optical emission linked to unobscured SF, coinciding with bright H$\alpha$ emission. To the northeast, a prominent tidal tail extends toward the C+D region. The spectra in Figure \ref{fig:Fig2} (green) indicate a superposition of structures between this tidal tail and the C+D region from the double peak emission lines.
Although the WFM does not resolve the individual sources in C+D visible in Figure \ref{fig:Fig1}, the brightest H$\alpha$ emission is found in the D region. To the northwest of this region, multiple sources in H$\alpha$ and the continuum are observed in the direction of \zwB. This region is characterized by the overlap of the three main structures—C+D, \zwA, and \zwB—referred to as the ``triple overlap region''. To the north direction is \zwB,  exhibiting a clear spiral structure, with several bright clumps in its arms and center, with an execs in MIR (Figure \ref{fig:Fig1}). 

At $\sim$ 7 kpc from the C+D region lies the E region, where bright filaments pointing towards region C are observed, possibly indicating a previous interaction. This region shows low H$\alpha$ emission and prominent H$\beta$ absorption line, indicative of a lower amount of ionized gas and a post-starburst (SB) episode \citep{Goto2007}.

The NFM's higher resolution reveals finer spatial details from the center of the systems and the interaction with their companions. In the NFM-South, several compact sources are found near the center, some of which have been identified by their MIR emission, possibly representing SF clumps with higher amounts of dust. In the northeast direction, the H${\alpha}$ and continuum emission highlight the emission from the C+D regions, connected by the tidal tail (represented by the white arrow), as the FoV of this data cube covers part of this region.

In the NFM-East, the structures from \zwA are also evident to the southwest, as its FoV includes parts of this galaxy. The increased spatial resolution reveals several compact sources, mostly detected by JWST/MIRI, suggesting a high dust amount. Their chaotic distribution suggests significant disruption from the merger, possibly involving the E region, as indicated by the connecting filaments in the WFM (Figure \ref{fig:Fig1}). Unfortunately, the NFM pointings do not cover the E region, preventing a more detailed analysis of this region. Finally, the NW NFM pointing contains multiple compact sources, that from its red excess in the MIR (Figure \ref{fig:Fig1}), suggest clumps of SF with higher amounts of dust.

\subsection{Kinematics}
\label{sec:Kinematics}

\begin{table*}[t]  
\centering
\caption{
Kinematic properties of the systems.
}  
\label{tab:kin_table}  
\resizebox{\textwidth}{!}{  
\begin{tabular}{l c c c c c}
\toprule
System & V$_{\text{sys}}$ & PA ($^{\circ}$) & $i$ ($^{\circ}$) & $R$ (kpc) & V$_{\text{max}}$ (km/s) \\ 
\midrule
\zwA    & -269 km/s  & 99$^{\circ}$ $\pm$ 24$^{\circ}$    & 40$^{\circ}$ $\pm$ 21$^{\circ}$  & 2  & 57 \\
\zwB    & -148 km/s  & 170$^{\circ}$ $\pm$ 5$^{\circ}$  & 65$^{\circ}$ $\pm$ 15$^{\circ}$        & 9  & 165 \\
D1     & -156 km/s  &       &            &    &     \\ 
\bottomrule
\end{tabular}
}

\vspace{5pt} % Espacio pequeño entre la tabla y las notas

\resizebox{\textwidth}{!}{%
\parbox{\textwidth}{%
\textsc{Notes.} Kinematic properties of the systems.  
(1) System name.  
(2) Systemic velocity (V$_{\text{sys}}$) in km/s, relative to z = 0.0365.  
(3) Position angle (PA) in degrees.  
(4) Inclination angle ($i$) in degrees.  
(5) Spatial extent of the system ($R$) in kiloparsecs (kpc).  
(6) Maximum velocity (V$_{\text{max}}$) in km$/$s.%
}%
}
\end{table*}

\begin{figure*}[!ht]
    \centering
    \includegraphics[width=1\linewidth]{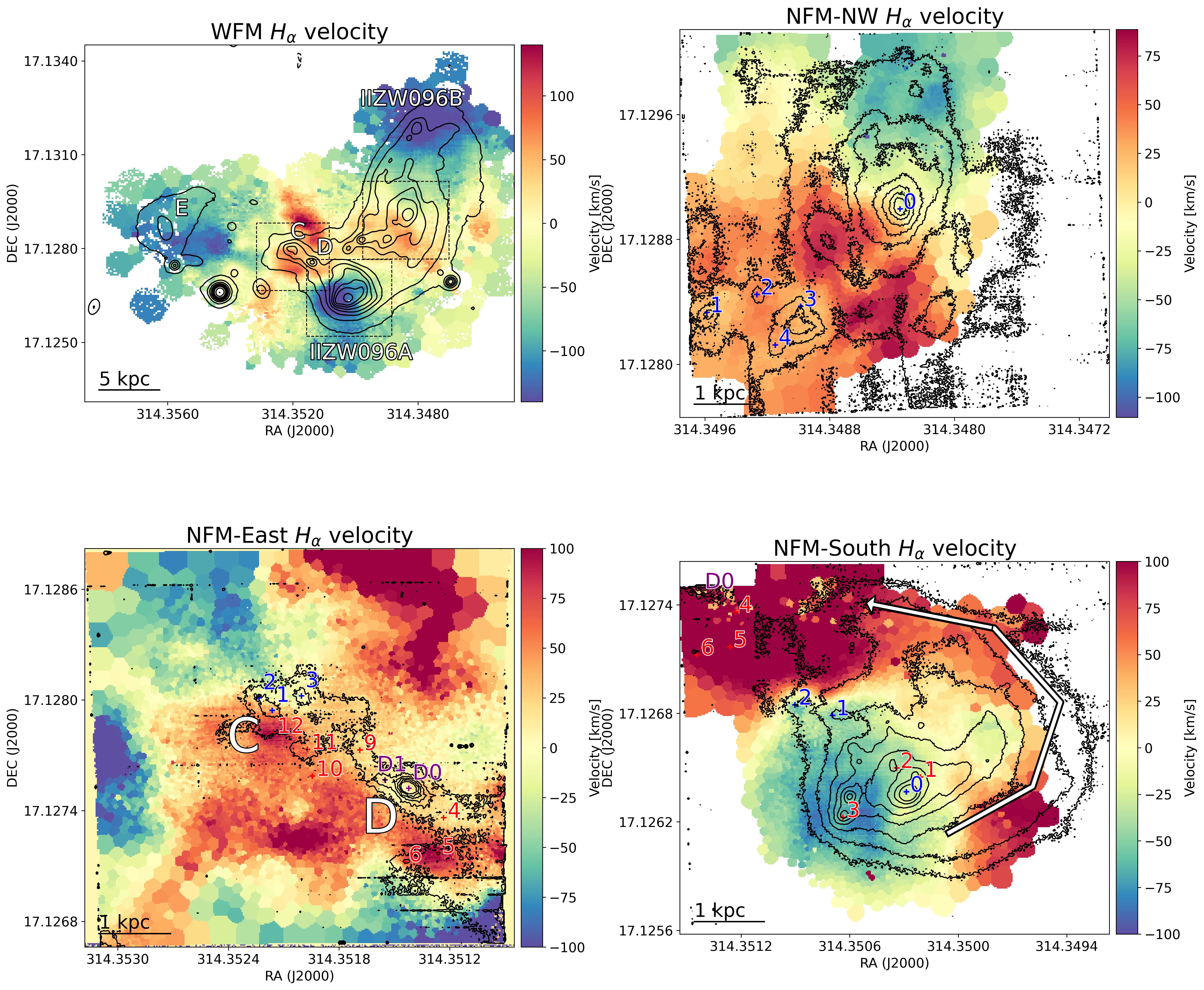}
    \caption{Velocity map extracted from the $H_{\alpha}$ emission line. The contours, regions, and sources are defined in Figure \ref{fig:mom0_halpha} .\textsl{Top left:} WFM map, with the velocity centered on \zwB and the FoV of the NFM pointings marked with dashed lines.\textsl{Top right:} NFM-South map, with the velocity centered on the systemic velocity of \zwA, highlighting an observed gradient along the major axis, with strong deviations near the edges.\textsl{Bottom left:} NFM-East map. Velocity is centered on D1, with no clear gradient visible, indicating relatively complex kinematics. \textsl{Bottom right:} NFM-NW map. Velocity is centered on galaxy nucleus B0, highlighting an observed gradient along the major axis.}
    \label{fig:mom1_halpha}
\end{figure*}

\begin{figure*}[!ht]
    \centering
    \includegraphics[width=1\linewidth]{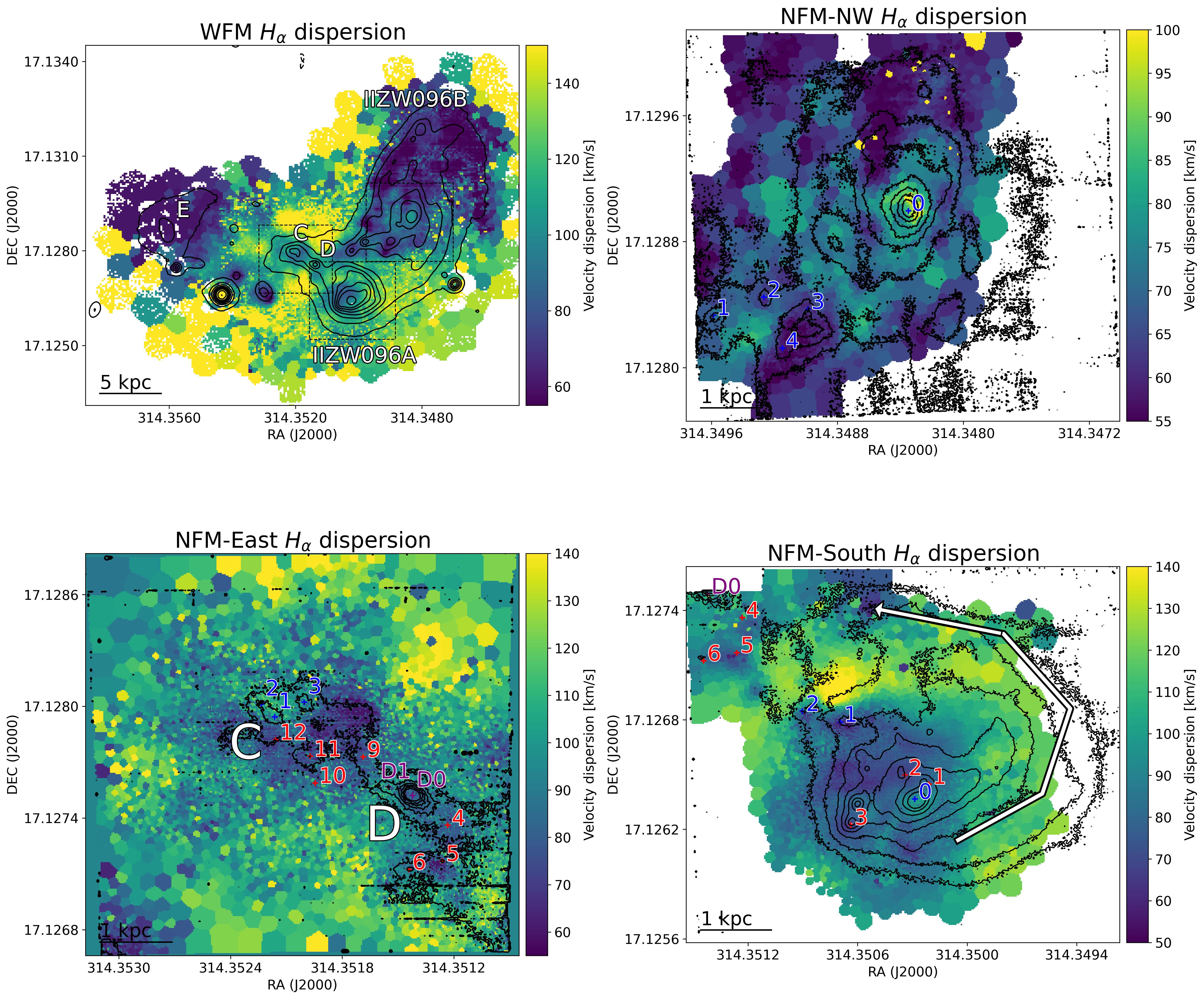}
    \caption{Velocity dispersion of the $H_{\alpha}$ emission line. The contours, regions, and sources are defined in Figure \ref{fig:mom0_halpha}. \textsl{Top left:} Velocity dispersion of the WFM cube with the NFM pointings marked with dashed rectangles. Regions with velocity dispersion $>$120 km$/$s exhibit double-peaked emission lines due to the overlap of structures at different velocities. \textsl{Top right:} NFM-South map, where, as in the WFM, the overlap with the C+D region and \zwB results in higher dispersion ($>$ 120 km/s). \textsl{Bottom left:}  NFM-East map, where the overlap with \zwA in the southwest direction results in double-peaked lines. In the north and south directions, similar profiles are observed. \textsl{Bottom right:} NFM-NW map, revealing higher dispersion in the center of \zwB.}
    \label{fig:mom2_halpha}
\end{figure*}

\begin{figure*}[!ht]
    \centering
    \includegraphics[width=1\linewidth]{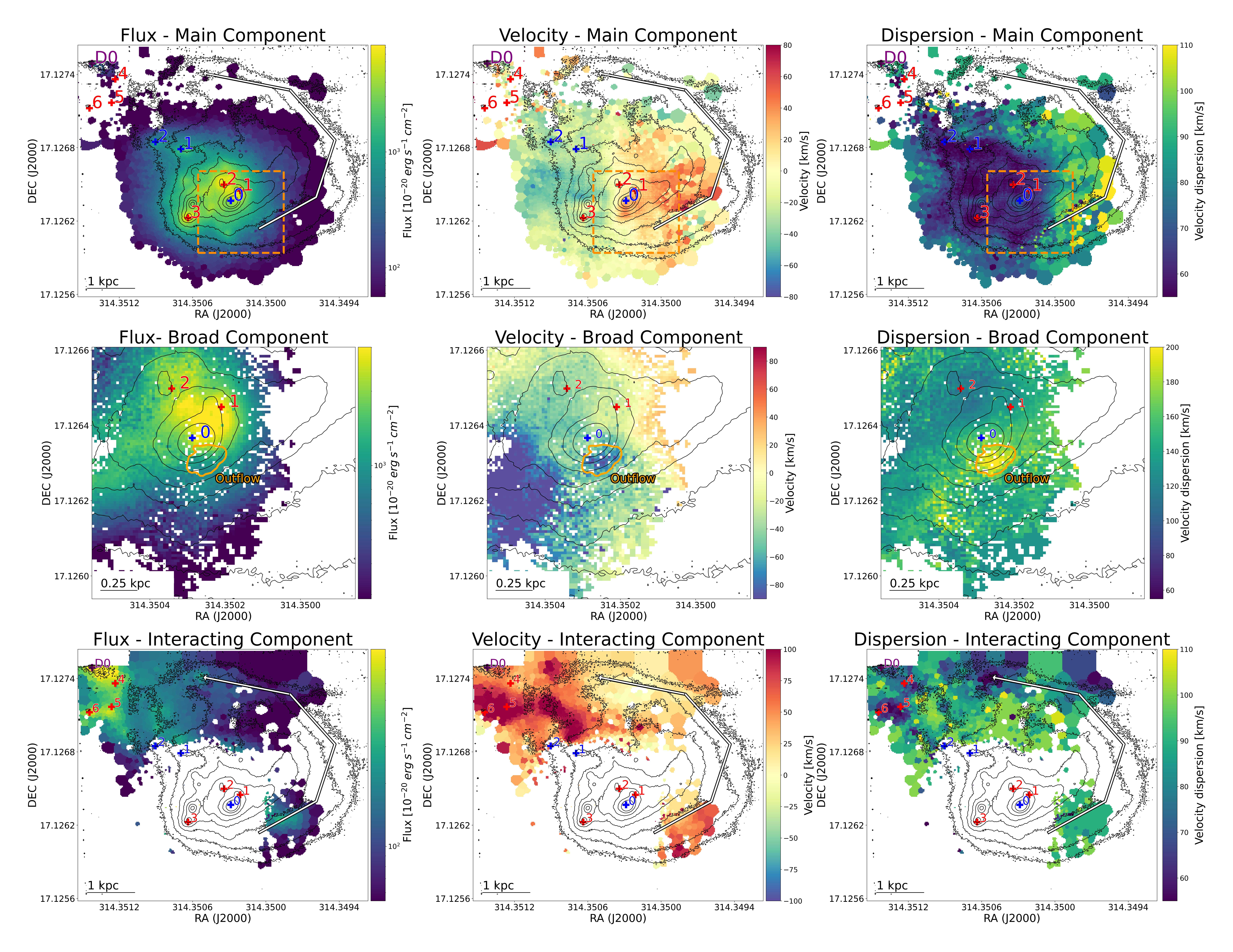}
    \caption{Maps of the flux, velocity, and velocity dispersion of the three different components from the $H_{\alpha}$ emission line in NFM-South. The contours, regions, and sources are defined in Figure \ref{fig:mom0_halpha}.\textsl{Top panels: } Maps of \zwA, referred to as the main component. The flux map reveals the same compact sources near the center and the tidal tail extending toward the C+D region. The velocity map shows a velocity gradient that deviates in the western regions with double-peaked emission lines and a redshifted component near the center. The velocity dispersion increases in the same direction as the double peaks blend due to limited spectral resolution. The dashed rectangles indicate zoomed-in views, as shown in the central panels. \textsl{Central panels:} Maps of the broad component, zoomed into the central part of the galaxy. The velocity maps reveal a gradient perpendicular to the main velocity gradient, accompanied by the high-velocity dispersion ($>$200 km$/$s) source PO. \textsl{Bottom panel: } Maps of the interacting component in the direction of the C+D region and the tidal tail extending toward \zwB. The map is centered on the $V_{sys}$ of D1. 
    }
    \label{fig:sep_comp_south}
\end{figure*}

\begin{figure*}[!ht]
    \centering
    \includegraphics[width=1\linewidth]{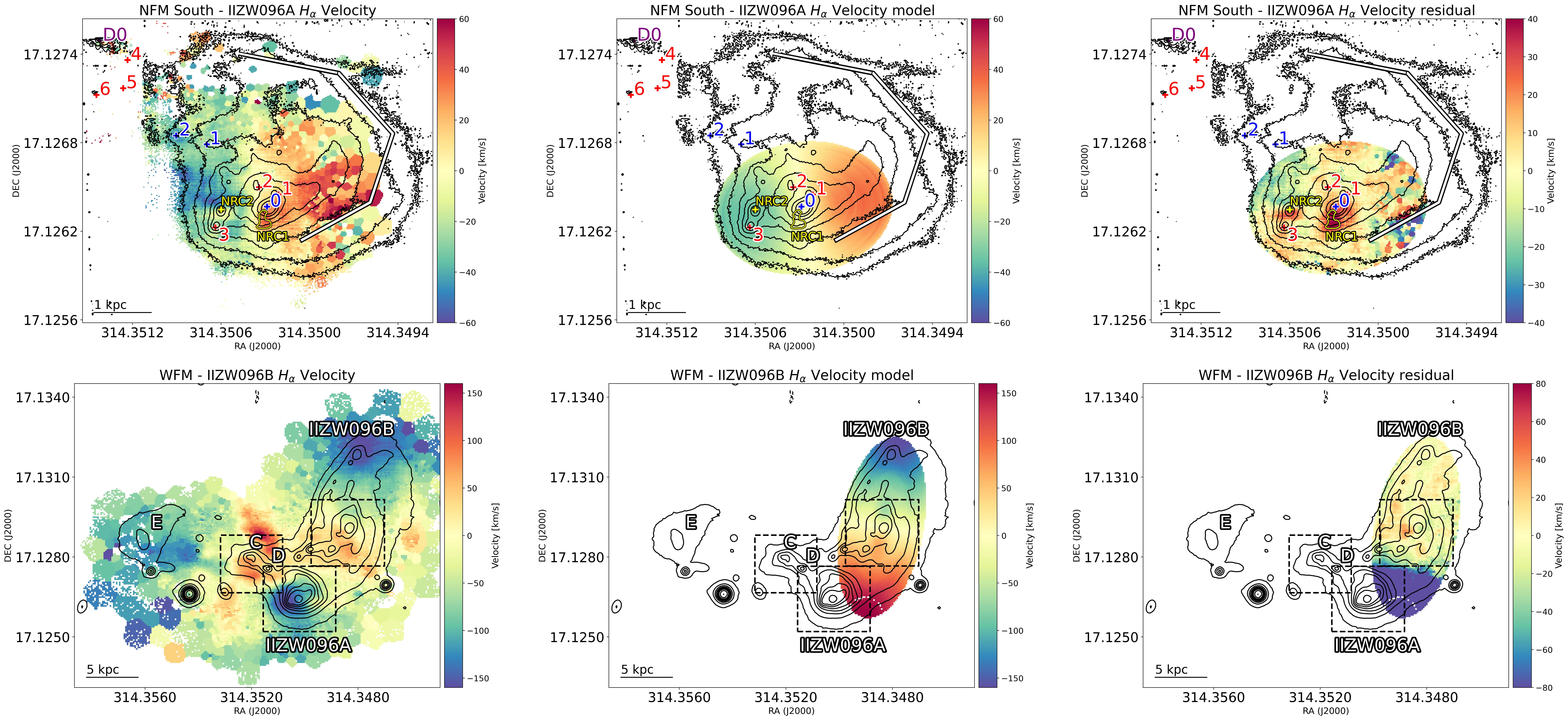}
    \caption{\texttt{Bbarolo} models for \zwA and \zwB. The contours, regions, and sources correspond to those already defined in Figure \ref{fig:mom0_halpha}. The left column presents the data, the central column presents the model, and the right the residuals (data-model). For \zwB, only the blueshifted side was used to generate the model.\textsl{Top-panels: } \texttt{Bbarolo} model of \zwA fitting the NFM-South data, using a PA = 92$^{\circ}$ and a $i$ = 40$^{\circ}$, with a maximum velocity of 57 km$/$s. \textsc{The residuals highlight the NRC1 in the center and the NRC2 in the north direction of ID3}. \textsl{Top-panels: } \texttt{Bbarolo} model for \zwB, using the WFM data, using a PA=170$^\circ$ and a $i$=65$^\circ$ with a maximum velocity of 165 km$/$s. The residuals suggest that \zwA is located in front of the redshifted region of this galaxy, with dust obscuring the redshifted side of \zwB. }
    \label{fig:bbarolo}
\end{figure*}

Fits to the detected emission lines were used to obtain the line-of-sight (LOS) velocity and velocity dispersion of the ionized gas. Figures \ref{fig:mom1_halpha} and \ref{fig:mom2_halpha} represent the velocity and velocity dispersion, respectively, of the total H$\alpha$ emission line fit, and Table \ref{tab:kin_table} contain the kinematic parameters from the different regions (see Section \ref{sec:kinematic_modelling} for more details). All velocity measurements are relative to z = 0.0365. 
\textsc{The instrumental spectral resolution of MUSE ranges from 72 to 46 km$/$s, from H$\beta$ to [S\,\textsc{II}]$\lambda$6731, respectively. Specifically for H$\alpha$, one of the key lines used in our kinematic analysis, the instrumental resolution is approximately 50 km$/$s. This value sets a lower limit on the velocity dispersions that can be reliably measured from the MUSE data.} \textsc{We find small variations in the velocity dispersion measurements in the overlapping regions of the South and East NFM pointings. These differences are smaller than $\sim$6 km/s, while the uncertainty in the velocity dispersion measurements is $\sim$10-15 km/s.}

In Figure \ref{fig:mom1_halpha}, the velocities in the WFM map are relative to the systemic velocity of \zwB center, -148 km$/$s, revealing a clear velocity gradient due to rotation. \zwA also shows a velocity gradient in the east and west directions, with significant deviations at the edges, where the double-peak emission lines appear, as shown in Figure \ref{fig:Fig2}. At the outer regions of \zwA, as the galaxy begins to fade, the velocity gradient of \zwB becomes apparent, which can be explained if \zwA is in front of \zwB, where dust is hindering its emission, and therefore, its velocity. A similar situation occurs between \zwA and the C+D region.
The C+D regions reveal complex kinematics, likely due to a more advanced merger stage. To the north of D, a redshifted stripe is observed, that appears to converge toward D. This could indicate material flowing towards this region. In the E region, a gradient is observed along a position angle (PA) of 165$^{\circ}$, but it is challenging to characterize due to the spatial resolution of the WFM map.

The NFM reveals more details about the kinematics. In the NFM-South, where the velocity map is relative to the systemic velocity of \zwA center (Table \ref{tab:kin_table}), as we approach the edges, in regions where double peaks are observed, the gradient deviates because the emission from the overlapping component begins to dominate.  In the NFM-NW, where the map is relative to the systemic velocity of \zwB center, the velocity gradient is even more pronounced. In the NFM-East, where the velocity map is relative to the velocity of D1 (-156 km$/$s), despite the increase in spatial resolution, no velocity gradient is observed, further highlighting the complexity of this region. Also, the redshifted stripe of $ \geq $120 km$/$s is observed more in detail, converting in what appears to be between D0 and D1.

The velocity dispersion map of the WFM, shown in Figure \ref{fig:mom2_halpha} reveals several regions with $\sigma$$>$120 km$/$s, like the superposition of \zwA with the C+D region and with \zwB, and to the north and south direction of the C+D region. These regions are characterized by double-peak line profiles, as shown in Figure \ref{fig:Fig2}, where the two Gaussian components exhibit similar $\sigma$. This type of profile is commonly observed in mergers \citep{Comerford2018}. However, the possibility of outflows can be ruled out due to their low-velocity dispersion and location far from the galaxy center. This suggests that their origin is more likely the result of the superposition of merging structures at different velocities.

The NFM observations show an increase in velocity dispersion in the NFM-NW. The center of \zwB exhibits values of $\sigma\sim$100 km$/$s, consistent with turbulence or shocks \textsc{ \citep[][ see Section \ref{sec:Ionization source} for further details]{Monreal-Ibero2006, Monreal-Ibero2011}}. The NFM-East region shows a region with $\sigma$$>$130 km$/$s between C1, C2, and C3, which are also characterized by double-peak emission lines. These features likely originate from structures with different velocities, possibly remnants of the interactions. The bright sources, such as D1 and D0, show values around $\sim$90 km$/$s, consistent with hosting shocks \textsc{\citep{Monreal-Ibero2006, Monreal-Ibero2011}}. The redshifted stripe of $\sim$ 120 km$/$s displays a velocity dispersion of $\sim$ 120 km$/$s, characterized by containing double peaks.

As observed from the NFM-South velocity dispersion map, \zwA contains multiple double-peak regions. Also, near A0, the center of the galaxy, a broad component is observed (Figure \ref{fig:Fig2}), so thanks to the increased spatial resolution of this observation, we can separate these components to analyze the kinematics of \zwA and better understand the origin of these complex line profiles. The criterion used to separate the emission from \zwA and the interacting components is based on the velocity continuity. Specifically, regions where double-peaked emission lines show abrupt velocity changes are identified as interacting structures. This method applies to velocities within the range of [-80, 80] km$/$s, which corresponds to the velocity of the blue-shifted side of the galaxy, less affected by interactions (centered on the systematic velocity of \zwA, Table \ref{tab:kin_table}). Outside this velocity range, components are classified as interacting structures. In the central region, all components with $\sigma$$>$120 km$/$s are classified as broad components, as the maximum velocity dispersion observed for the narrow components near the center does not exceed this value.

Figure \ref{fig:sep_comp_south} presents the different components' flux, velocity, and velocity dispersion maps. The first row shows the maps of \zwA, with a dashed orange rectangle, indicating the zoomed-in region of the second row, which corresponds to the broad component. Both are relative to the systemic velocity of \zwA. The third row represents the interacting component relative to the systemic velocity of D1 (Table \ref{tab:kin_table}), given that part of these structures belong to the C+D region. 
In the upper row, the velocity gradient of \zwA is clear, with a PA of $\approx$100$^\circ$. At the galaxy's edges, the velocity gradient becomes less distinct as the overlapping structures dominate the emission. The sources A1 and A2 are likely part of \zwA, as their systematic velocities are closer to this galaxy. At $\sim$2 kpc to the northwest, the tidal tail's velocity dispersion ($\sigma$) reaches values $>$110 km$/$s, likely due to unresolved double-peaked lines in large Voronoi-binned spaxels. In the galaxy's center, a redshifted component of $\sim$50 km$/$s is detected, pointing in the south direction. Due to its low-velocity dispersion ($\sigma \sim 80$ km$/$s), the possibility of it being an outflow is low and is more likely to be a structure related to the merger.
The middle row reveals a high-dispersion structure ($\sigma > 200$\,km$/$s) with slightly blueshifted velocities, reaching up to 90 km$/$s. While this structure does not exhibit particularly high fluxes or velocities, its velocity dispersion is consistent with the presence of an outflow\textsc{; we will refer to this feature as PO (potential outflow). Its location is defined based on criteria from a sample of local weak starburst outflows \citep{Couto2021}, identifying this biconical structure with velocities between $-60$ and $-90$ km$/$s and velocity dispersion exceeding 180 km$/$s. We apply this threshold as a mask for future calculations and to highlight this region in Figure~\ref{fig:sep_comp_south}.}
The bottom row presents the interacting structures. In the east direction, sources ID4, ID5, and ID6 are identified, likely part of the D region due to their velocities being closer to the systemic velocity of D1. Conversely, to the west of \zwA, a newly identified structure appears, which, when compared to Figure \ref{fig:Fig1}, seems to be a tidal bridge resulting from an interaction, pointing in the direction of \zwB.

%%%%
%%BBArolo
%%%%%
\subsubsection{Kinematic modeling}
\label{sec:kinematic_modelling}
We use the \texttt{Bbarolo} software \citep{DiTeodoro2015} on its 2D version mode to model the velocity fields, focusing exclusively on the kinematics. This code reconstructs velocity fields with independent tilted rings, making it possible to fit one side of the system (e.g., the approaching or receding side of the galaxy) and symmetrize the results on the opposite side (see, e.g., \citet{delMoralCastro2019}). We fitted \zwA and \zwB, where potential rotating components are observed. The same process was attempted for the D0 and D1 regions, but in these cases, a reliable fit could not be obtained due to the small size of the source, which prevented the construction of robust models. All parameters obtained are summarized in Table \ref{tab:kin_table}

For \zwB, we considered only the blueshifted side since \zwA obscures the redshifted part. An attempt was made to use both sides of the galaxy, yielding the same position angle (PA) and inclination ($i$) as when using only the blueshifted side. However, the extension of the rotating disk could not be determined due to its relative position concerning \zwA. Meanwhile, for \zwA, both sides were considered for the model. The WFM data was used for \zwB because its larger FoV allows us to capture the full extent of the galaxy. This was not necessary for \zwA, as the NFM already covers it completely. %The BBarolo algorithm is highly sensitive to the initial parameter guesses. Therefore, multiple models with varying $i$ and the PA were tested to compare their reduced $\chi^{2}$ values. 

The initial parameter to determine is the systemic velocity, v\(_{\text{sys}} \), as this will locate the kinematic center of the system. For \zwA, we allowed this parameter to vary freely and observed that it converged to \(-269 \, \text{km$/$s}\), which differs from the value measured at the center, or A0 (\(-239 \, \text{km$/$s}\)). This discrepancy arises due to the presence of the redshifted velocity structure of \(50 \, \text{km$/$s}\) in the southern direction, which was previously identified. This structure is located near the center, and its emission obscures the true systemic velocity. By allowing this parameter to vary freely in \texttt{Bbarolo}, the value of \(-269 \, \text{km/s}\) was the one that converged, enabling symmetrical velocities around a stable PA. With the v$_{sys}$ obtained, the model converge to a PA = 99$^\circ$ $\pm$ 24$^\circ$, and $i$ = 40$^\circ$ $\pm$ 21$^\circ$. For \zwB, the values converge to an v$sys$ = 148km$/$s, PA = 170$^\circ$ $\pm$ 5$^\circ$, and $i$ = 65$^\circ$ $\pm$ \textsc{15}$^\circ$, all summarized in Table \ref{tab:kin_table} . Figure \ref{fig:bbarolo} shows the resulting model and residual maps. The top row shows \zwA, indicating that the rotation extends up to 2 kpc, with a maximum velocity of 57 km$/$s. From the residual, the galaxy is consistent with rotation. However, the regions in the outer parts of the galaxy exhibit high residuals, where the influence of interactions is more prominent, causing the rotational motion to no longer be followed. Additionally, two prominent structures deviate from the rotation.
\textsc{The most prominent structure is the redshifted feature near the center, named NRC1 (Non-Rotating Component 1), which is likely related to a structure produced by the merger due to its low dispersion ($\sigma \; = 80$ km$/$s). Based on its velocity dispersion and a velocity range between 40 and 50 km$/$s, we use this criterion to define the region. The second structure is the compact region located in the northern direction of ID3, named NRC2 (Non Rotating Component 2), which, compared to the surrounding clumps, is not bright in H$\alpha$ (Figure \ref{fig:mom0_halpha}), possibly being a dense region of stars moving at different velocity compared to the rotation of the galaxy. NRC1 and NRC2 are shown in Figure \ref{fig:bbarolo}.} In the bottom panels, \zwB exhibits a rotating distribution with no prominent residuals, up to approximately 9 kpc,  reaching a velocity of $\sim$ 165 km$/$s. The redshifted region, to the south, is not able to be fully fitted due to the influence of \zwA.

%Viviane suggest to objectivetly assess the fit, maybe a chisquare

\subsection{Ionization sources}
\label{sec:Ionization source}

\begin{figure*}[!ht]
    \centering
    \includegraphics[width=1\linewidth]{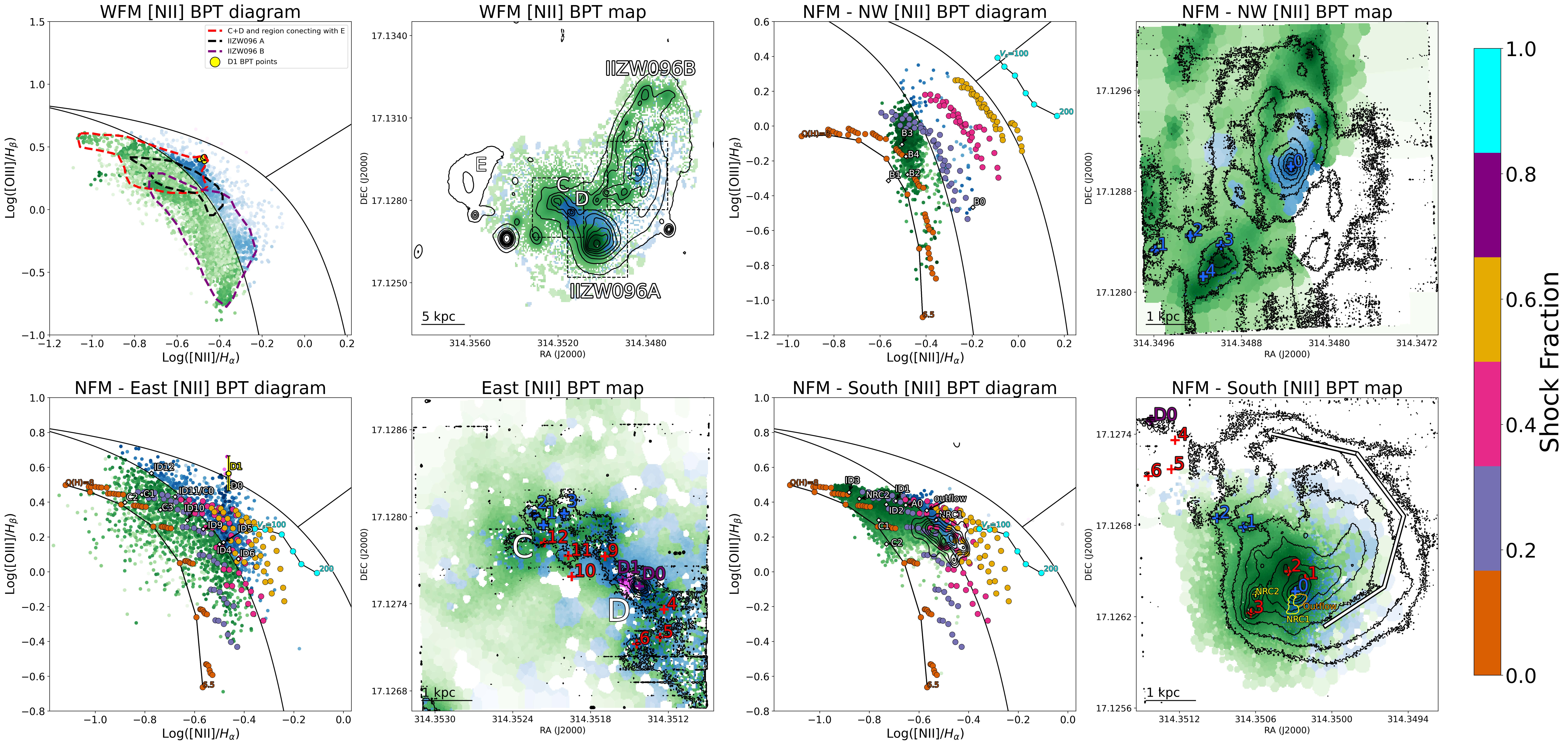}
    \caption{[N II]$\lambda$6584/H${\alpha}$ diagnostic diagrams \citep{Baldwin1981} of the WFM and NFM. The contours, regions, and sources were already defined in Figure \ref{fig:mom0_halpha} \textsc{, in addition to the sources NRC1, NRC2 and PO presented in Figure \ref{fig:bbarolo} and \ref{fig:sep_comp_south}}. The green color represents SF, blue represents composite, magenta represents Seyfert, and gray represents LINER, with each color weighted by the H$\alpha$ flux. \textsc{The HII and shock models from \citet{Farage2010} and \citet{Rich2010} are shown as points outlined in black. These represent a shock contribution ranging from 0\% (orange points connected by black lines) to 100\% (cyan points connected by black lines). The intermediate points correspond to shock contributions of 5\%, 20\%, 40\%, and 60\%}. \textsc{In the diagrams, the point sources presented in the maps are shown according to their respective mean line ratios. For the NFM-South diagram, the black contours represent the distribution of the emission lines of the tidal tail of \zwA}. The yellow points in the WFM map represent the line ratios of D1, the same as for the NFM-East, but with the latter being resolved. The error bars for the D1 point are 0.01 and 0.1  in the x and y axis. In the bottom-left panel, the NFM-East diagnostic diagram, the yellow contours indicate the location of the ALMA band 3 continuum, highlighting the concentrated emission in D1. The theoretical limit for ionization in a pure SF scenario is indicated by the black line from \citet{Kewley2001}. In the [N II] diagram, the curve below this limit represents the classical separation between SF and Seyfert, as reported by \citet{Kauffmann2003}. The regime between these curves, defined as composite, marks the transition between high-ionization SF and Seyfert. The line separating the LINER and Seyfert classifications is taken from \citet{Schawinski2007} for the [N II] diagram and from \citet{Kewley2008} for the [O I] and [S II] diagrams.}
    \label{fig:Bpt_nii}
\end{figure*}

\begin{figure*}[!ht]
    \centering
    \includegraphics[width=1\linewidth]{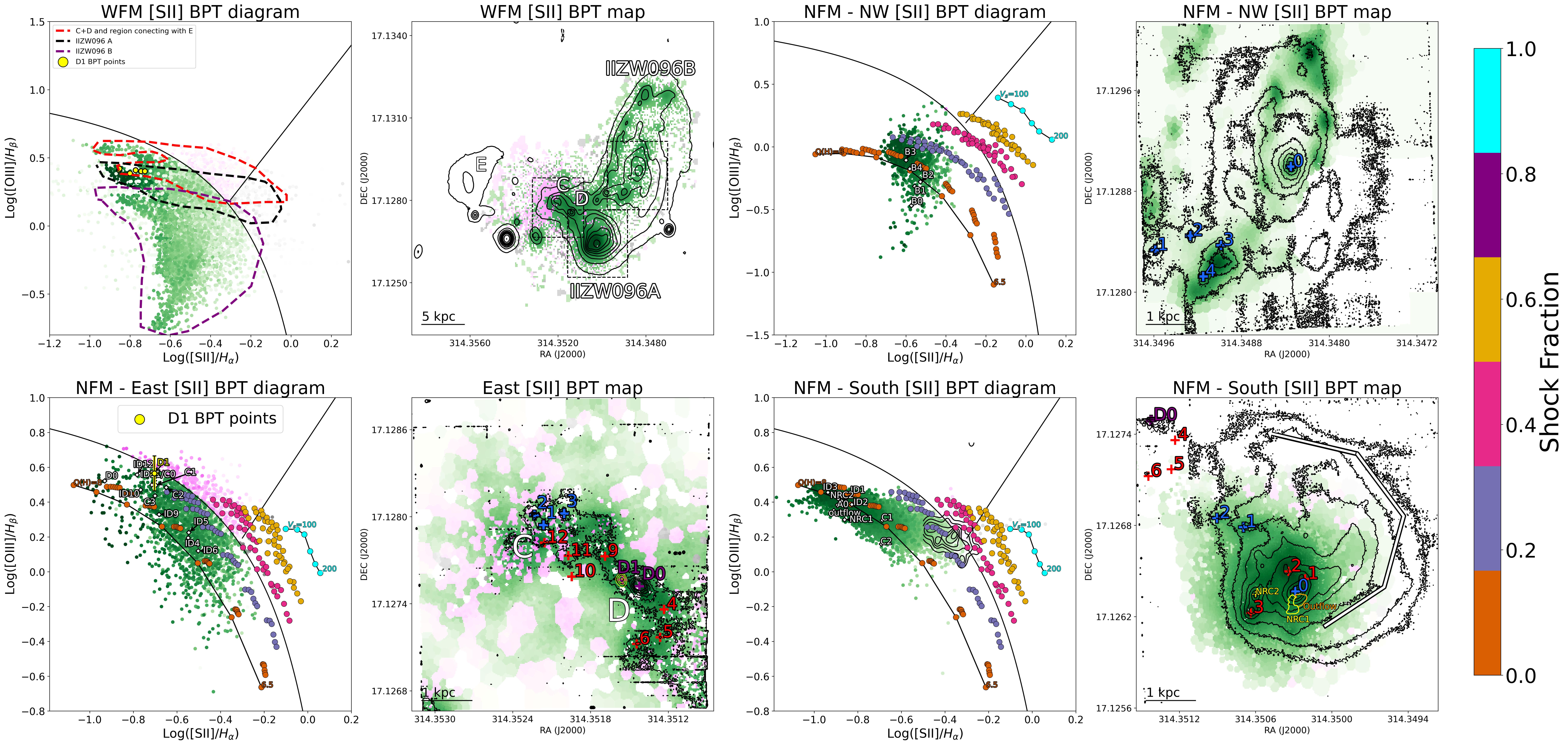}
    \caption{Same as Figure \ref{fig:Bpt_nii}, for the [S II]($\lambda$6716 + $\lambda$6731)/$H_{\alpha}$ diagnostic diagram \citep{Veilleux1987}. The green color represents SF, magenta represents Seyfert, and gray represents LINER. The error bars for the D1 point are 0.02 and 0.1  in the x and y axis. }
    \label{fig:Bpt_sii}
\end{figure*}

\begin{figure*}[!ht]
    \centering
    \includegraphics[width=1\linewidth]{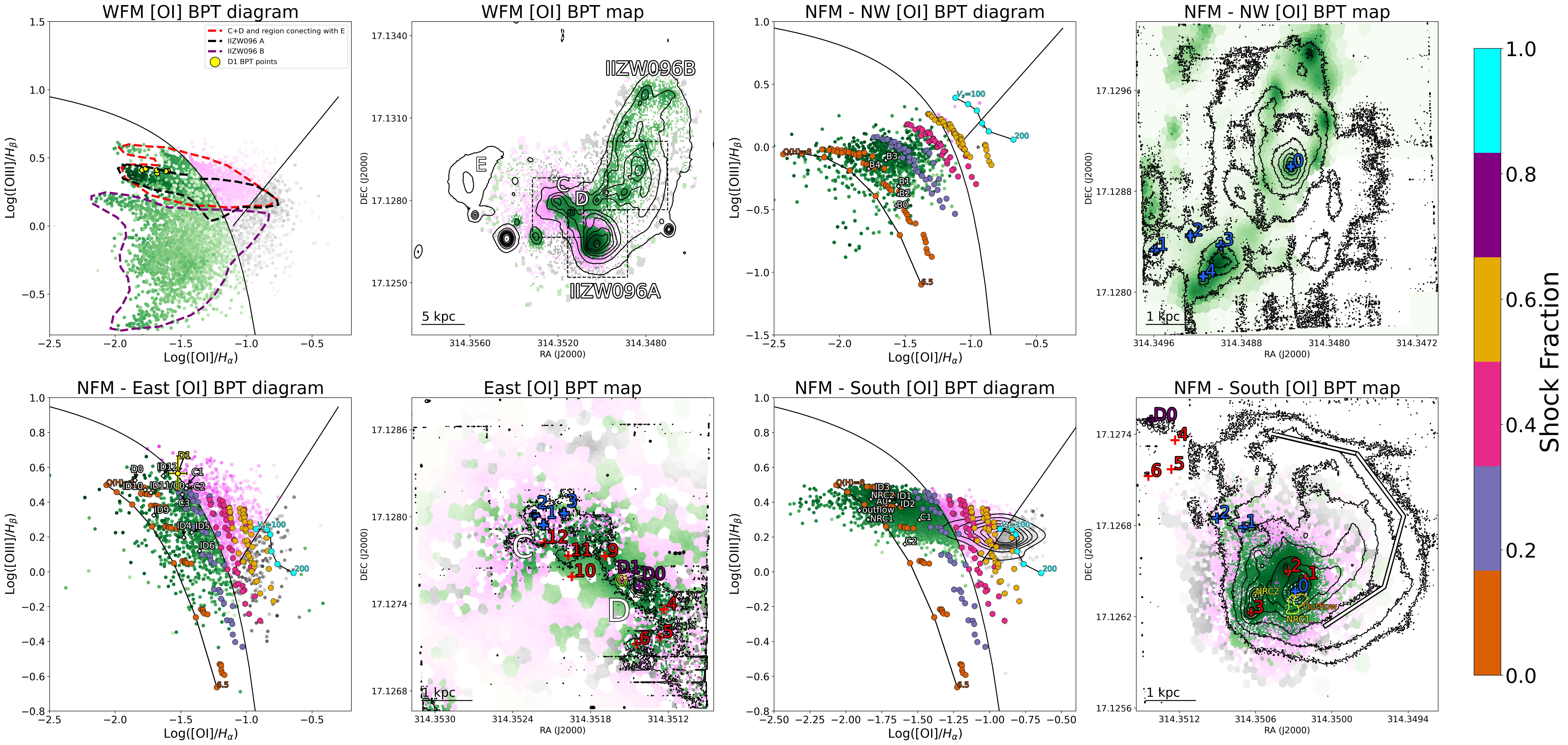}
    \caption{Same as Figure \ref{fig:Bpt_nii}, showing in this case the [O I]$\lambda$6300/$H_{\alpha}$ diagnostic diagram \citep{Veilleux1987}. The green color represents SF, magenta represents Seyfert, and gray represents LINER. The error bars for the D1 point are 0.07 and 0.1  in the x and y axis. }
    \label{fig:Bpt_oi}
\end{figure*}

The IFU data allows to classify the different regions acording to their principal ionization mechanism. In the optical range, the most common method for classification is the optical diagnostic diagram \cite{Baldwin1981, Veilleux1987}. These diagrams utilize a combination of emission lines to classify sources into categories such as SF, Seyfert, Low-Ionization Nuclear Emission-Line Regions (LINERs), and Composite systems, which are powered by more than one ionization mechanism. To ensure reliable measurements, we apply Voronoi binning based on the H$\beta$ emission line, requiring a minimum S/N $=$ 3, and excluding spaxels with S/N $< 1$.

We employ three diagnostic diagrams for classification: 
\begin{itemize}
    \item The [NII] diagram, using the line ratios $\log([\text{O III}]\,\lambda5007/\text{H}\beta)$ versus $\log([\text{N II}]\,\lambda6583/\text{H}\alpha)$,
    \item The [SII] diagram, using the line ratios $\log([\text{O III}]\,\lambda5007/\text{H}\beta)$ versus $\log([\text{S II}]\,\lambda(6717+6731)/\text{H}\alpha)$,
    \item The [OI] diagram, using the line ratios $\log([\text{O III}]\,\lambda5007/\text{H}\beta)$ versus $\log([\text{O I}]\,\lambda6300/\text{H}\alpha)$.
\end{itemize}
\textsc{The flux line ratios maps are presented in Appendix \ref{sec:Appendix Line Ratio} for comparison.} \textsc{Extinction has a negligible effect on the BPT diagram because it involves ratios of emission lines that are very close in wavelength, minimizing differential extinction. } Figures \ref{fig:Bpt_nii}, \ref{fig:Bpt_sii}, and \ref{fig:Bpt_oi} show these diagrams for the WFM and NFM data. We adopt the classification schemes from \citet{Kewley2001, Kauffmann2003, Schawinski2007, Kewley2008} to distinguish between Seyfert (magenta), SF (green), LINER (grey), and composite (blue) regions. For the WFM data, we use the flux from the full Gaussian fit, while for the NFM-South, only the emission identified as originating from \zwA in Section \ref{sec:Kinematics} is used for the line flux ratios. In the other NFM regions, a single fitted component is used for the analysis. In both the diagnostic diagrams and the maps, the data is weighted by H$\alpha$ flux to make it easier to identify the spatial regions in the diagrams.

\textsc{For the WFM diagrams and maps, we have marked the regions of \zwA\ with black dashed lines, those of \zwB\ with purple dashed lines, and the regions C+D and the one connecting to E with red dashed lines. The E region has not been included due to the low S/N of the fainter lines required to produce each diagnostic diagram. In the NFM diagrams, we plot a mixing sequence from purely HII photoionized regions to completely shock-excited models. These shock models were introduced by \citet{Farage2010} and \citet{Rich2010}, who used the MAPPINGS code \citep{Allen2008}. The velocities generated by these models were based on the observations of merging LIRGs in \citet{Rich2011}, which, compared with \zw, show similar values. The HII region models were generated using \texttt{Starburst99} \citep{Leitherer1999}. These regions are plotted with varying ionization parameters $Q(H)$, the flux of ionizing photons divided by the hydrogen atomic density, ranging from $\log Q(H)$=8-6.5, consistent with typical values for HII regions \citep{Dopita2000}. The shock models are shown for velocities ranging from [200 - 100]~$\mathrm{km/s}$. In addition to the models, the identified sources, which were presented in Figure \ref{fig:Fig1}, are shown in the diagnostic diagrams, each represented by the mean of their line ratios. In addition, the mean line ratios of the sources NRC1, NRC2, and PO from  NFM-South are also included. For all the sources represented in the line ratio diagrams, in the participating emission lines, we have previously subtracted the estimated contribution due to the emission from the surrounding galaxy. This was done by computing the emission-line fluxes in nearby regions around each analyzed source, which were then appropriately scaled by the corresponding areas and subtracted from the source's flux in each emission line, thus allowing us to isolate the emission from each analyzed source.}

In the WFM diagrams, SF dominates the ionization across all three diagrams, particularly in \zwB. However, in the prominent tidal tail of \zwA, which extends toward the C+D region, the ionization is classified as composite in the [N II] diagram, SF in the [S II] diagram, and SF and Seyfert in the [O I] diagram. This shift is consistent with expectations for merging systems, likely driven by shocks from SB activity or gas interactions \textsc{\citep{Monreal-Ibero2006, Monreal-Ibero2011, Medling2021}}. The C+D region exhibits a combination of SF and Seyfert classifications in the [S II] and [O I] diagrams. At the same time, it falls within the SF and composite categories in the [N II] diagram. [O I] and [S II] are lower ionization potential (IP) lines, making them more sensitive to phenomena such as merger-induced shocks or winds from the starburst. These processes can significantly influence the diagnostic diagrams, so the increase in ionization is likely related to the region's highly disturbed morphology and kinematics, suggesting a past interaction. The yellow points in the WFM map represent the line ratios of D1, where the spatial region was selected based on NFM-East observations. However, due to the spatial resolution of the WFM and the proximity of D0, this emission is a combination of both sources. As a result, its ionization classification is under composite, but this is likely severely affected by the contamination from the surrounding sources near D1.

\textsc{In all three NFM diagrams, there is clear evidence of strong star formation activity, as indicated by the prominent H$\alpha$ emission. This makes it possible to rule out more evolved low-mass stars, such as AGB stars, as the main ionizing sources. These stars are typically associated with older stellar populations, which, due to their harder spectra, tend to fall within the LINER classification \citep{Belfiore2016,Belfiore2022}. Therefore, in the regions covered by our three NFM pointings, younger stellar populations, shocks, or possibly AGN feedback are more plausible ionizing sources.} 

\textsc{In the NFM-South diagrams, SF is the dominant ionization mechanism. Except for PO, all regions in the diagram fall under this classification across the three diagnostic diagrams. Compared with the HII + shock models, the shock contribution ranges from 0 to 20\%, with HII regions characterized by $\log(Q(H))$ between 7.25 and 8. Examining the sources A0, ID1, ID3, and NRC2, we find that they are consistent with the highest values of $Q(H)$. Their spectra exhibit a prominent ``blue bump'' around 4650\,\AA, a feature characteristic of Wolf-Rayet (WR) stars. WR stars represent a short-lived evolutionary phase \citep{Meynet2005} of the most massive stars ($\geq 25 \; M_{\odot}$), occurring after they leave the main sequence, and are characterized by their emission of very energetic photons, consistent with the higher values of $Q(H)$. (For reference, a map of the blue bump distribution is shown in Appendix~\ref{sec:Appendix blue bump}.) Based on the study of the merging LIRG Antennae \citep{Weilbacher2018}, leaking Lyman-continuum (LyC) photons ($\lambda < 912$\,\AA) from HII regions have been identified as a viable ionization mechanism. In this study, the line ratio of [O\,III]~$\lambda5007$ / [S\,II]~$\lambda\lambda6716,6731$ was used as a sensitive diagnostic of the ionization parameter and can indicate density-bounded regions that allow LyC escape. In such cases, the low-ionization zone is reduced or absent, leading to high [O\,III]/[S\,II] ratios. In the central region of the Antennae Galaxy, values of [O\,III]/[S\,II]~$> 2.5$ were found to be strong indicators of LyC leakage, despite no general trend across all HII regions. Using this value as an indicator, the regions A0, ID1, ID3, and NRC2 exhibit ratios higher than this limit, where the WR stars could provide the energetic photons responsible for the LyC leakage. }

\textsc{The NRC1 region, due to its higher $\log([\text{N\,II}]\,\lambda6583/\text{H}\alpha)$ ratio and distinct kinematics compared to the rotation pattern, is a candidate for being ionized by alternative mechanisms such as shocks. Meanwhile, the PO region shows the highest shock contributions in the system, ranging from 5 to 40\%. Given its high velocity dispersion ($\sigma \sim 200$\,km/s), an outflow origin cannot be excluded. The black contours in all three diagrams represent the line ratios of the tidal tail in the C+D direction. These are consistent with a higher shock fraction, ranging from 40 to 100\%. This agrees with its classification as Seyfert and LINER in the [S\,II] and [O\,I] diagrams, and as composite in the [N\,II] diagram, where an increase along the x-axis of all three diagrams indicates a higher flux of photons with lower energy ionizating the gas, caused by shocks \citep{Monreal-Ibero2006, Monreal-Ibero2011}. In the NFM-NW diagrams, the central region B0 shows a maximum shock contribution of 20\% in the [N\,II] diagnostic, explaining its classification as composite in this diagram. This slight increase in ionization, possibly related to turbulence or shocks, is also reflected in the increase in velocity dispersion, as shown in Figure~\ref{fig:mom2_halpha}. The best-fitting HII region model corresponds to $\log(Q(H))$ values between 6.75 and 7 for this source. Meanwhile, the other sources are consistent with HII regions and no shock contribution, with $\log(Q(H))$ values from 7 to 7.25. In the NFM-East pointing, the error bars were derived from the fit uncertainties in D1, which are assumed to be representative of measurements in this region. The point sources show a higher contribution from shocks, reaching up to 40\% with $\log(Q(H))$ values ranging from 7 to 8, consistent with young SP influenced by stronger shock contributions. The second-brightest source, D0, is consistent with an HII region of $\log(Q(H)) = 7.75$ in all three diagrams, in agreement with previous studies identifying it as a powerful star-forming region \citep{Donnan2024}. The regions with higher values along the x-axis in the three diagrams exhibit stronger shock contributions, from 40 to 100\%, and are located in the exterior parts of regions C and D (see Appendix~\ref{sec:Appendix Line Ratio} for reference). These are classified as either Seyfert, LINER, or Composite. Based on their complex morphology and kinematics, it is reasonable to suggest that this system has been significantly affected by interactions, leading to enhanced ionization due to shocks. Some regions in the C region show higher values of $\log([\text{O\,III}]\,\lambda5007/\text{H}\beta)$, between 0.5 and 0.7, and lower values of $\log([\text{N\,II}]\,\lambda6583/\text{H}\alpha)$, between $-0.9$ and $-0.8$ (see Appendix~\ref{sec:Appendix Line Ratio} for reference). Compared to the SB line ratios from the Antennae galaxy in \cite{Gunawardhana2020}, these regions are consistent with HII regions ionized by a starburst with a high value of $Q(H)$ and relatively low gas metallicity. These regions also exhibit [O\,III]/[S\,II]~$>$2.5, suggesting the presence of LyC leaking due to this younger SP activity. The yellow point corresponds to the line ratios obtained by integrating the spectrum of D1 with a radius of 0.23\arcsec, based on the measurements from \citet{Inami2022}. In the [N\,II] diagram, this point is classified as Seyfert, although it lies near the classification boundary, especially considering the error bars. In the [S\,II] and [O\,I] diagrams, the point is classified as SF, but if we consider the error bars, the Seyfert classification cannot be ruled out. In addition, the models cannot fully reproduce the observed line ratios, raising the possibility of an additional ionizing source in this region. This will be further discussed in Section~\ref{sec:Discussion}.}

In all three diagnostic diagrams, the yellow contours on the map correspond to the continuum emission from ALMA Band 3, with D1 standing out as the brightest source, exhibiting very compact emission. This suggests that D1 is a compact source emitting in the millimeter wavelength, driving increased ionization and approaching the threshold for AGN classification in the optical diagnostic diagrams.

\subsection{Extinction}
\label{sec:Extinction}

\begin{figure*}[!ht]
    \centering
    \includegraphics[width=1\linewidth]{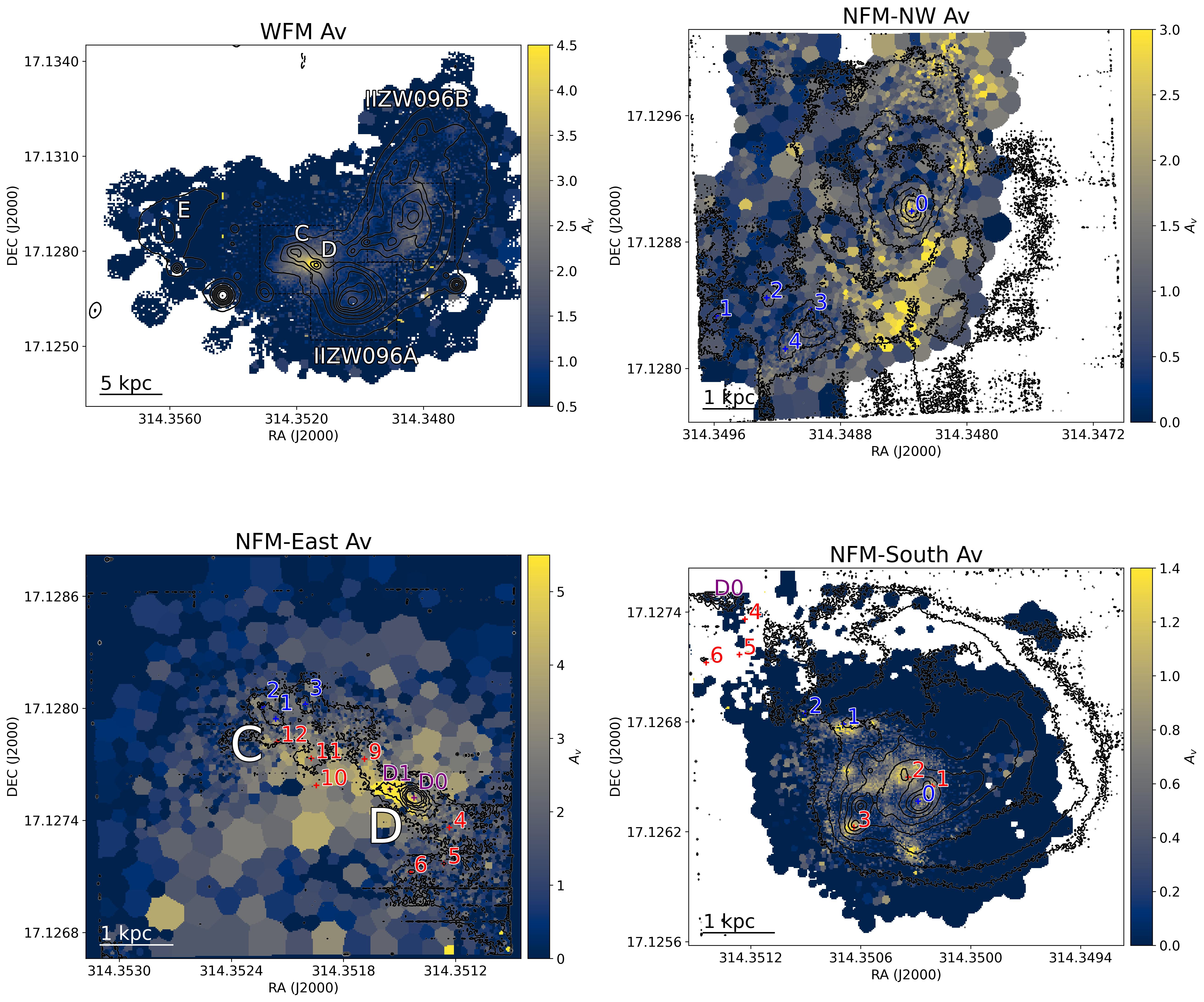}
    \caption{A$_{H\alpha}$ map of the WFM and NFM, derived from the observed H$_{\alpha}/$H$_{\beta}$ ratio. The contours, regions, and sources are defined in Figure \ref{fig:mom0_halpha}.\textsl{Top left:} WFM map, where the D region presents the highest A$_{H\alpha}$ values.\textsl{Top right:} NFM-South map. The compact sources detected by JWST/MIRI contain the highest A$_{H\alpha}$ values. The dashed white area represents the high A$_{H\alpha}$ bent structure in the direction of \zwB. \textsl{Bottom left:} NFM-East map. The entire C+D region is characterized by high A$_{H\alpha}$ values, especially in D1. \textsl{Bottom right:} NFM-NW map. The compact sources, identified by their strong H$\alpha$ emission, exhibit the highest A$_{H\alpha}$ values.
    }
    \label{fig:extinction}
\end{figure*}

High optical extinction is common among major galaxy mergers, where high concentrations of gas and dust are present, particularly in the case of (U)LIRGs \citep{Veilleux2009}. The optical extinction can be parametrized through A$_{H\alpha}$, which can be estimated from the Balmer decrement, $H_{\alpha}/H_{\beta}$. We employed the \citet{Calzetti2000} attenuation law for a galactic diffuse ISM, with $R_V$=3.12, and an intrinsic Balmer decrement H${\alpha}$/H${\beta}$=2.86, assuming a Case B Recombination, for an electron temperature $T$= 10$^4$K, a standard choice in SF galaxies \citep{Osterbrock1989}. Figure \ref{fig:extinction} presents the derived extinction values from the WFM and NFM data cubes.

The WFM map highlights the obscured C+D region, reaching values up to A$_{H\alpha} \sim 4.5$ in D, emphasizing the large amounts of dust in this region. \zwA has an A$_{H\alpha}\sim 1.5$ at its center, while \zwB shows similar values at the center of the galaxies and in the SF clumps in its arms. The E region does not show high extinction values, indicating a relatively lower dust amount.

The NFM-South shows that the bright clump in H$\alpha$ from Figure \ref{fig:mom0_halpha} has A$_{H\alpha}$=1.4, suggesting that these are regions of intense SF with a large amount of dust. To the south of the nucleus, an area with similar A${_H\alpha}$ values is observed, marked with dash lines, spatially coinciding with the structure previously identified as the redshifted component near the center in the kinematic modeling (Figure \ref{fig:bbarolo}). This structure exhibits a velocity dispersion of $\sigma \sim 80$ km$/$s. It extends up to 0.7 kpc, bending towards \zwB, to the region of double-peaked lines, and may represent a structure produced by an interaction.

The NFM-NW pointing shows high extinction values primarily in three regions: the center of the galaxy and two sources located north and south of the nucleus, with A$_{H\alpha}\simeq 3$. These regions also coincide with bright emission in H$\alpha$, similar to \zwA: SF clumps associated with a large amount of dust. In NFM-East, D1 is responsible for the highest  A${_H\alpha}$, with a peak value of 5.4. However, the entire C+D region is characterized by a high extinction of A$_{H\alpha}$$\sim$3 due to large amounts of dust distributed around the whole area. 

Using the Balmer decrement, we can correct the H${\alpha}$ flux to derive the intrinsic spatially-resolved SFR of the galaxies, a crucial parameter to characterize them physically. It could not be done for the C+D region, as it has not yet been determined whether all the emissions originate from SF. The main ionization mechanism of the other galaxies has been identified as SF in section \ref{sec:Ionization source}, so we can assume that the majority of the H${\alpha}$ emission is produced by this process. Adopting the relation from \citet{Kennicutt1998}, the SFR for \zwA is SFR$_{H\alpha}$=12 M$_{\odot}$ yr$^{-1}$, while for \zwB, it is SFR$_{H\alpha}$=18 M$_{\odot}$ yr$^{-1}$. \citet{DS2017} report the total SFR of both galaxies, obtained through X-ray observations, as SFR$_{H\alpha}$=30.6 M$_{\odot}$ yr$^{-1}$, which is consistent with the sum of the values we have derived for both galaxies.
As for the E region, an SFR$_{H\alpha}$=0.4 was obtained, consistent with its faint H$\alpha$ emission.

\section{Analysis of Chandra X-ray Data}
\label{sec:Chandra analysis}

\begin{figure*}[!htb]
    \centering
    \includegraphics[width=1\linewidth]{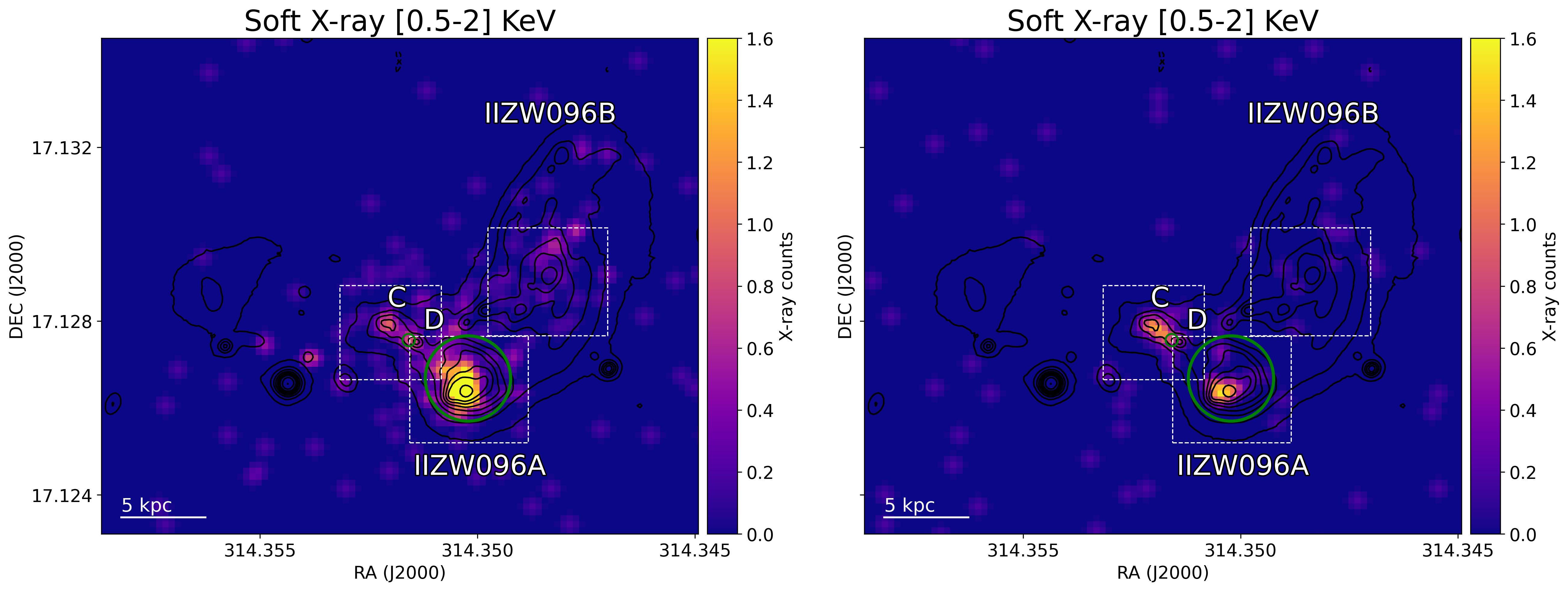}
    \caption{Smoothed X-ray images using a Gaussian filter with a 0.5\arcsec$\times$0.5\arcsec radius from the soft, 0.5-2 keV, and hard, 2-8 keV, bands. The contours, regions, and sources are the same as defined in Figure \ref{fig:mom0_halpha}. The green circles represent the apertures used to calculate the HR in D1 and \zwA.}
    \label{fig:chandra}
\end{figure*}

\subsection{Data description}
\label{sec:data description chandra}

The X-ray data were taken with Chandra Advanced CCD Imaging Spectrometer \citep[ACIS;][]{Garmire2003}  from \citet{Iwasawa2011}, on 2007-09-10, with an exposure time of 14.57 ks. The data reduction was done following standard procedures, using CIAO v.4.15 and the Chandra Calibration Database, CALDB v.4.10.7. Data were reprocessed with the \texttt{chandra\_repro} task to generate the final event table. In this analysis, we focus on the comparison between the soft([0.5-2]KeV) and hard([2-8]KeV) X-ray data. The \texttt{dmcopy} task was used to separate these energy ranges and measure the total count in each aperture.

\subsection{X-ray analysis}
\label{sec:X-ray analysis}

In Figure \ref{fig:chandra}, we present the soft ([0.5-2] keV) and hard ([2-8] keV) X-ray emission detected by Chandra. The images have been smoothed using the \texttt{csmooth} CIAO task, using a Gaussian filter with a circular radius of 0.5\arcsec, for better visualization. \zwA exhibits the strongest emission in the soft X-ray, with emission from the galaxy itself and extending into the tidal tail in the direction of the C+D region. The C+D region exhibits dominant hard X-ray emission, with its peak slightly offset from the D1 source ($\sim$ 1\arcsec). Nonetheless, D1, localized by the green circle, exhibits significant emission in the hard X-ray regime. In the soft band, to the south of the filament connecting the east region with E, there are three faint sources, two of which coincide with the optical continuum. \citet{Iwasawa2011} describe these sources as possible remnants of a past interaction, which could be connected with the E region due to its vicinity. 

The Hardness Ratio (HR), defined as $HR = (H - S)/(H + S)$, where $H$ and $S$ are the background-corrected counts in the hard and soft bands, respectively, was used to characterize the spectral shape of the X-ray emission. This color index is useful when there are insufficient counts to generate a statistically meaningful spectral fit, as for D1. This value was calculated using the Bayesian Estimator of Hardness Ratio \citep[BEHR;][]{Park2006}. This code estimates the HR and its uncertainties, focusing on the low-counts Poisson regime using the source and background counts in the two energy bands. The counts in the different bands were obtained with the \texttt{dmstat} task from CIAO. For \zwA we use an aperture centered on the galaxy (RA: \( 20^{\text{h}} 57^{\text{m}} 24.03^{\text{s}} \), DEC: \( +17^{\circ} 07' 36.07'' \)), with a radius of 3.53\arcsec, to enclose the galaxy and its most prominent tidal tails, represented by the green circle. As for D1, we use an aperture, represented by the green circle, of 0.5\arcsec radius. The HR of  \zwA is -0.56 $\pm$ 0.08, having an excess of soft x-rays, dominated by low energy photons possibly originated in SF and shock processes. \citet{Iwasawa2011} obtained an HR = -0.57 $\pm$ 0.07 for the entire system, indicating that \textsc{in \zwA} predominantly dominates the X-ray emission. The HR of D1 is 0.1$\pm$0.3. The error bars in this case are large due to the low number of counts of the source, unlike \zwA. However, D1 is still consistent with a hard source. Compared to \citet{Iwasawa2011}, the C+D region has an HR = 0.05 $\pm$ 0.14, which is still consistent with an obscured AGN. However, this X-ray color is contaminated by the surrounding sources near D1.

\section{Discussion}
\label{sec:Discussion}

\begin{table*}[t]  
\centering
\caption{
Gas-phase metallicity of the systems.
}  
\label{tab:metal_table}  
\resizebox{\textwidth}{!}{  
\begin{tabular}{l c c c c c}
\toprule
System & 12+log(O/H)  & Center (RA$^{\circ}$, Dec$^{\circ}$) & $i$ ($^{\circ}$)  & $a$ (\si{\arcsecond}) & $b$ (\si{\arcsecond}) \\ 
\midrule
\zwA    & 8.53 $\pm$ 0.14 & 314.356, 17.129 & 0   & 6               & 6              \\ 
\zwA tidal tail & 8.58 $\pm$ 0.11 & 314.349, 17.127 & 55  & 11              & 3              \\ 
\zwB    & 8.65 $\pm$ 0.15 & 314.348, 17.130 & 165 & 11              & 6              \\ 
E      & 8.46 $\pm$ 0.16 & 314.356, 17.129 & 45  & 6               & 5              \\ 
C+D    & 8.52 $\pm$ 0.2 & 314.352, 17.128 & 65  & 4               & 3              \\ 
\bottomrule
\end{tabular}
}
\vspace{5pt} % Espacio pequeño entre la tabla y las notas
\resizebox{\textwidth}{!}{%
\parbox{\textwidth}{%
\textsc{Notes.} 12 + log(O/H) of the systems in \zw, with the respective elliptical apertures to obtain this value.  
(1) System name.  
(2) Gas-phase metallicity expressed as 12+log(O/H) with the N2 method, using \citet{Curti2019} calibration.  
(3) Center of the aperture to obtain the gas metallicity (RA, Dec).  
(4) Inclination angle ($i$) in degrees.  
(5) Semi-major axis ($a$) in arcseconds.  
(6) Semi-minor axis ($b$) in arcseconds.%
}%
}
\end{table*}

\begin{figure*}[!htb]
    \centering
    \includegraphics[width=1\linewidth]{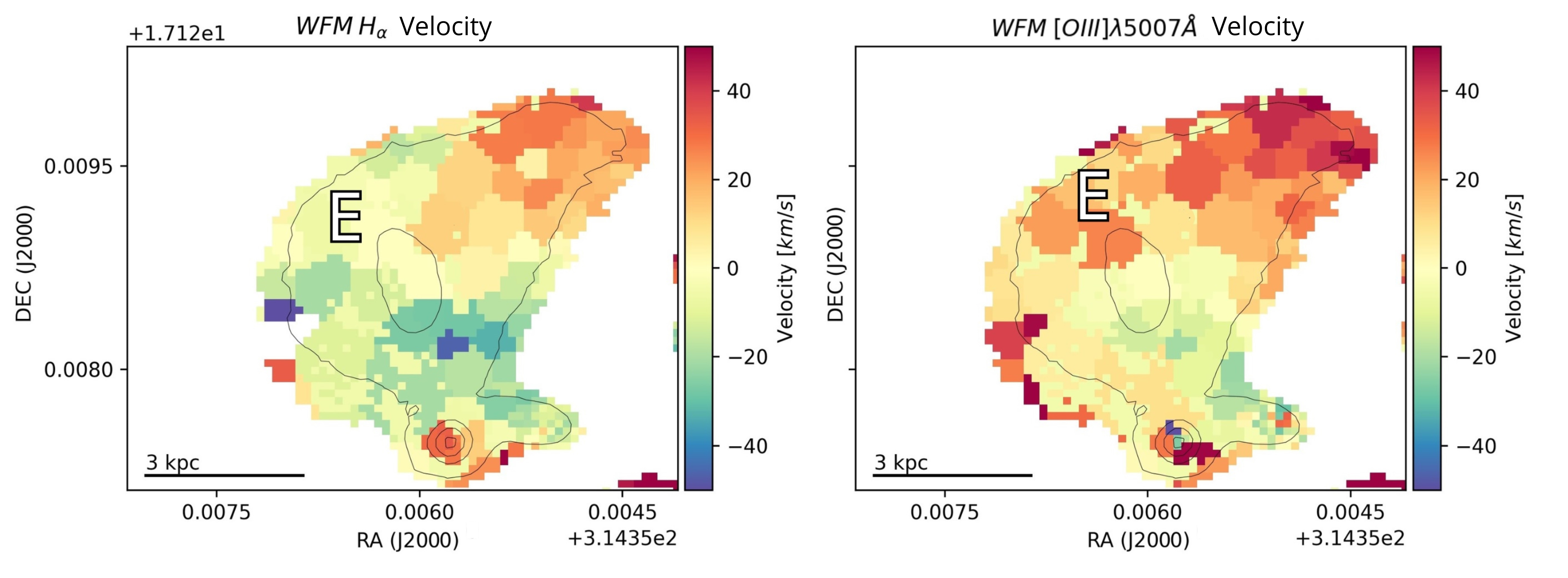}
    \caption{H$\alpha$ and [O III]$\lambda$5007 Velocity maps for the E region. The contours are defined in Figure \ref{fig:mom0_halpha}, while the scale bars represent a physical length of 3 kpcs.}
    \label{fig:kin_e}
\end{figure*}

\subsection{Characterizing Ionization and Interaction}
\label{sec:Characterizing Ionization and Interaction}

In this section, we describe the interactions and ionization around the system, starting from the west side of the merger, composed by \zwA and \zwB, and moving in the east direction, which contains the C+D and E regions.

% Describing the interaction and ionization in the "eastern" side of the interaction

% First describing the interaction of the south galaxy, that is in front of NW and C+D

\zwA shows evidence of two prominent structures: the tidal tail in the C+D direction and the other in the direction of \zwB, \textsc{the NRC1. For the NRC1 region, we gained insight into the interaction by combining kinematics, ionization, and extinction in the NFM-South region. From the \texttt{Bbarolo} model (Figure~\ref{fig:bbarolo}), it is clear that this central structure does not follow the galaxy’s rotation pattern, with velocities reaching up to 50 km$/$s. Its distinct kinematics, the higher $\log([\text{N\,II}]\,\lambda6583/\text{H}\alpha)$ ratio compared to other compact sources in the galaxy, and its point source-like shape, all suggest that it may be ionized by the remnant of a star, such as a planetary nebula (PNe) or a supernova remnant (SNR). In the original BPT diagram \citep{Baldwin1981}, PNe typically show $\log([\text{N\,II}]\,\lambda6583/\text{H}\alpha)$ between 0.0 and 0.5, and $\log([\text{O\,I}]\,\lambda6300/\text{H}\alpha)$ between 0.0 and 0.5, both higher than the values observed in this source. Additionally, they exhibit $\log([\text{O\,III}]\,\lambda5007/\text{H}\beta) \geq 1$, which clearly rules out this scenario. As for the SNR, a commonly used criterion to classify SNRs, particularly ones with low shock velocities, is $\log([\text{O\,I}]\,\lambda6300/\text{H}\alpha)$$>$-1.76 \citep{Kopsacheili2020}. The NRC1 region shows values consistent with this criterion. Unfortunately, these models are based on the contribution of shocks to the ionization of emission lines, while in merging systems, other mechanisms, such as tidal interactions, can also produce shocks. Using these diagnostics alone to distinguish SNRs in such environments is hence unreliable. Therefore, for NRC1, an SNR origin cannot be excluded, nor a contribution from shocks induced by the interaction with the companion galaxy.}
 But this structure shows A$_{H\alpha}$ values up to 1.4 (Figure \ref{fig:extinction}), bending in the direction of \zwB, represented by the dashed lines. This phenomenon becomes evident when compared with Figure \ref{fig:Fig1}a. This redshifted structure, characterized by high $A_V$, likely represents a tidal tail produced by an interaction, possibly a bent arm containing a high SF clump with a large amount of dust (as indicated by its bright H$\alpha$ emission and elevated A$_{H\alpha}$)—a recurring feature in this system. Its velocity dispersion of $\sim$ 80 km$/$s\textsc{, significantly higher than the instrumental resolution,} is consistent with a structure generated by interaction \textsc{\citep{Monreal-Ibero2006, Monreal-Ibero2011}}, where shocks induced by this event lead to increased ionization, explaining its composite classification in Figure \ref{fig:Bpt_nii}.

% Ionization of the tidal tail
For the tidal tail in the direction of C+D, an increase in ionization has been identified, classified as Composite in the [NII] diagram or Seyfert and SF in the [SII] and [OI] diagrams (Figures \ref{fig:Bpt_nii}, \ref{fig:Bpt_sii}, \ref{fig:Bpt_oi}). These last two optical diagrams are sensitive to low-ionization events due to their lower IP, indicating that this ionization is likely not caused by a Seyfert but by other phenomena, such as \textsc{shocks, winds from SB or photoionization from more evolved stars}. \textsc{The line ratios, shown by the black contours in the diagnostic diagrams (Figures~\ref{fig:Bpt_nii}, \ref{fig:Bpt_sii}, and \ref{fig:Bpt_oi}), indicate a significant shock contribution, ranging from 40 to 100\%. This is consistent with the increase along the x-axis, where emission lines with lower ionization potential are typically enhanced in the presence of shocks \citep{Monreal-Ibero2006, Monreal-Ibero2011}. From the X-ray observations, \zwA shows a hardness ratio (HR) of $-0.56 \pm 0.08$, characterized by soft emission originating from the tidal tail in the C+D direction. Combining the optical and X-ray data, this region appears to be ionized by shocks, which compress and heat the gas, producing collisional ionization that enhances the emission of low-ionization species. This suggests an increase in the number of ionizing photons with relatively low energies. In addition, as shown in Figure~\ref{fig:Fig2}, the tidal tail exhibits a prominent H$\beta$ absorption line — a signature of post-starburst (post-SB) events \citep{Goto2007}. This region, which resembles diffuse ionized gas, is consistent with photoionization from both young massive stars and possibly by hot, evolved low-mass stars,  typical of post-SB populations, as observed in galaxies with recent star formation \citep{Belfiore2016, Belfiore2022}. Such conditions can produce line ratios characteristic of LINER, Seyfert, or composite classifications in diagnostic diagrams. The observed increase in ionization likely results from a combination of shocks and post-starburst activity, suggesting that \zwA is already being affected by the interaction. In contrast, \zwB appears relatively undisturbed, retaining its spiral arms and showing largely stable rotation and morphology. In the optical diagnostic diagrams, both the ionization from the sources and the gas remain at levels not typically associated with shocks or winds from intense star formation triggered by interactions. This is further supported by the low shock contribution indicated by the models, reinforcing the interpretation that \zwB is less affected by the merger.}

\textsc{In the center of \zwA, we have successfully identified the emission from PO, with velocities between $-60$ and $-90$ km$/$s and a velocity dispersion greater than 180 km$/$s. For outflows, the blueshifted velocities are easier to observe, as the obscuration by the galaxy can hide the redshifted counterpart of these winds. Compared to a sample of weakly ionized outflows from local star-forming galaxies analyzed by \cite{Couto2021}, these kinematic values fall within the range of such weak outflows. To compare their luminosities, we measure in this region \(L_{\text{H}\alpha} = 4 \times 10^{32} \, \text{W}\), which places it at the lower end of the outflow luminosities. This suggests that, if this is indeed an outflow, it is very weak, even compared to those in star-forming galaxies. This interpretation is consistent with its line ratios, which match models with only a 5 to 40\% shock contribution, supporting the idea of a relatively mild interaction with the surrounding gas.} Another possibility for the origin of this structure is that it could be part of the bent structure NRC1, but with different motions compared to it. Both possible origins can explain the increased ionization to composite in the center, either due to winds from high SF or by the interaction of structures due to the merger. \textsc{As for the compact sources A0, ID1, ID3, and NRC2, we have successfully identified ionization from WR stars and also found evidence suggesting LyC leakage, which could explain their similarities with HII region models featuring higher $Q(H)$ values. In contrast, the remaining sources, such as C1, C2, and ID2, are better matched by HII models with lower $Q(H)$.}

%ouu
%Esto solo lo puedo poner si hablo del SP, pero seria en el NFM de South
%Figure \ref{fig:age_wfm} shows that the tidal tail has an SP in the range between 0-2 Gyr. This provides information about the history of this interaction. The H$\alpha$ emission is not prominent in the tidal tail, indicating that there is not an SF rate currently, but it must have been high in the past, generating the young stars. A tentative scenario is that the shocks produced a high and fast SF, resulting in that SP, and now this activity has ceased, with no prominent H$\alpha$ emission and only the residual stars. This also gives an upper limit for this interaction, being no older than 2 Gyr.

% Now that I have been talking about the interaction with the NW, I start to talk about this galaxy
% Describing the interaction of the NW galaxy

%This structure is also detected by the predominant metallicity of -1.71 [M/H] (Figure \ref{fig:metall_wfm}).

%%%%%%%%%%%%%%%%%%%%%%%%%%%%

% Now that I have described the two less perturbed galaxies, I start to talk about the kinematics, their modeling, and mass ratio, explaining at the end that this is a major merger

%The west side of the interaction, the NW and South galaxy is less affected by the merger, deduced by its morphology and especially by the kinematics. With BBarolo, we modeled the rotation of the galaxy to analyze the residuals and obtain the total mass of the galaxies.

The kinematic modeling of \zwB and \zwA presented in Figure \ref{fig:bbarolo} provides valuable insight into their rotation and interaction. The upper panel in Figure \ref{fig:bbarolo} shows that \zwA is well described by a rotating model with a PA = 99$^\circ$ $\pm$ 24$^\circ$ and an $i$ of 40$^\circ$ $\pm$ 21$^\circ$, extending up to 2 kpc. The residuals highlight the previously discussed NRC1, suggesting its possible origin as a tidal tail. To the north of the ID3 source, \textsc{the NRC2 does not follow the galaxy's rotation, possibly representing a dense region of star formation (due to its location in the diagnostic diagram) that does not follow the galaxy's rotation pattern}. Similarly, \zwB is successfully modeled as a rotating disk with a PA = 170$^\circ$ $\pm$ 5$^\circ$ and an $i$ of 65$^\circ$ $\pm$ 15$^\circ$, extending up to 9 kpc, and its residuals do not reveal any prominent structures, supporting the argument that it is less affected by the merger.

To estimate the magnitude of the merger between these two galaxies, we compare the stellar masses of these two galaxies. Using the 3.5~$\mu$m method along with JWST/NIRCam F356W data, we derive approximate mass estimates, utilizing the kinematics to separate the components and determine the photometric apertures. These values provide a rough idea of the merger's category rather than precise measurements and help us assess the system's interaction. We use the \citet{Wen2013} correlation, specifically the equation for late-type galaxies, since both galaxies exhibit signatures of spiral arms, to derive the stellar mass:

\begin{equation}
\begin{split}
    \log_{10}\left ( \frac{M_{*}}{M_{\odot}} \right ) = (0.679 \pm 0.002) + \\
    (1.033 \pm 0.001) \times \log_{10}\left ( \frac{\nu L_{\nu}(3.4 \si{\micro\metre})}{L_{\odot}} \right )
\end{split}
\end{equation}

The 3.3 $\mu$m PAH feature falls within this filter. To correct for its contribution, we use the spectrum of \zw\ from the AKARI Infrared Camera \citep[IRC;][]{Onaka2007}. The spectrum is obtained from the entire system, meaning that the PAH contribution to the JWST/NIRCam F356W filter represents an upper limit. By masking the PAH emission line, we fit the continuum with a polynomial, selecting the lowest possible degree that minimizes the $\chi^2_{\text{reduced}}$. Considering the filter's response, we obtained a maximum PAH contribution of 25\% to the band. Including this contribution, we estimated stellar masses of $2.2  \pm 0.4\times 10^{10}\: M_{\odot}$ for \zwA and $4.5\pm 0.8 \times 10^{10}\: M_{\odot} $ for \zwB. While these values are approximate, they provide a useful indication of the interaction's magnitude, suggesting that this is a major merger, as its stellar mass ratio is 1.2-2.6. After characterizing \zwA and \zwB, we now move on to the east side of the merger, focusing on the regions C+D and E.

%fff

%evidence from past interaction

%\begin{figure*}[!htb]
    %\centering
    %\includegraphics[width=1\linewidth]{Figures/MS.jpg}
    %\caption{ SFR vs. Stellar Mass of the four sources in \zw, represented in black, and a sample of local (U)LIRGs from \cite{Eser2014} in blue. The dashed lines show the MS for $z \sim 0$ from \cite{Elbaz2007}. \zwA and \zwB are located with the local (U)LIRGs, while D1, due to its compactness and high SFR, is far from the local (U)LIRGs. However, these values should be taken as upper limits since the IR contribution of the AGN has not been accounted for. The E region is below the MS, classified as a post-MS system, based on its prominent H$\beta$ absorption features.
%}
%    \label{fig:ms}
%\end{figure*}

%% Starting up with the triple overlap region

On the east side of the system, the first prominent structure is the triple overlap region, combining the influence from the three main components: \zwA, \zwB, and C+D. \zwA is connected by a prominent tidal tail, while \zwB is linked by a structure with multiple bright clumps that, combined with the classification as SF in the optical diagnostics, highlights the presence of high SF activity.

% Now to the east region

\textsc{Further to the east lie the C+D and E regions, which appear to be the most disrupted by the merger, as evidenced by their disturbed morphology, complex kinematics, and line ratios in the diagnostic diagrams. As discussed in Section~\ref{sec:Ionization source}, the C regions shows regions suggesting the presence of LyC leaking, due to young SP. Also, the outskirts of regions C and D exhibit high shock contributions, ranging from 40 to 100\%, consistent with their classification as Seyfert and Composite. This interpretation aligns with the [Mg\,\textsc{IV}] analysis by \citet{Pereira-Santaella2024}, who also concluded that shocks play a dominant role in these regions. The compact sources within this area show elevated $Q(H)$ values, between 7 and 8, along with significant shock contributions. This suggests intense star formation activity coupled with shocks, likely driven by the interaction between regions E and C+D. Additionally, the velocity dispersion map (Figure~\ref{fig:mom2_halpha}) reveals multiple areas with double-peaked line profiles and total velocity dispersions exceeding 120 km$/$s. These features indicate the presence of kinematically distinct components along the same line of sight, likely the result of a past interaction, providing further evidence of the violent nature of this event.}
There is, however, some uncertainty regarding the E region. Is it a remnant of the C+D region or a separate system? \textsc{To address this, we analyze the kinematics of the E region to search for signatures such as rotation, which would support the galaxy scenario. We also examine the gas metallicity and compare it with that of the other system to check for any differences, further supporting the idea that it is a separate galaxy.} In Figure \ref{fig:kin_e}, we present the kinematics of the E region based on the [O III]$\lambda$5007 and H$\alpha$ lines. We observe a velocity gradient in both maps, with a consistent PA of approximately 165$^\circ$, suggesting rotational motion. We further applied the mass-metallicity relation (MZR) to assess whether the E region aligns with the scaling relation. The MZR, which describes the increase of metallicity with stellar mass due to processes regulating gas accretion, star formation, and outflows, uses the oxygen abundance relative to hydrogen ($12 + \log(\mathrm{O/H})$) as a proxy for gas metallicity. \textsc{This merger contains multiple diffuse ionized gas (DIG) regions, such as the E region and inner regions in \zwA and \zwB , which behave differently from HII regions, due to the lower surface brightness, differing physical conditions, and contamination by multiple ionization sources \citep{Dale2009, Belfiore2016}. Hence, standard gas metallicity estimators based on well-behaved HII regions, such as the N2 diagnostic ([N\,\textsc{II}]$\lambda$6583/H$\alpha$), are highly uncertain. Therefore, the method proposed by \citet{Pilyugin2016} was used, based on the line ratios of [N II] $\lambda$6549 + $\lambda$6583 $/$ H$\beta$ and [S II] $\lambda$6717 + $\lambda$6730 $/$  H$\beta$, which shows good consistency with abundances derived from auroral line temperature measurements}. 

\textsc{In Table \ref{tab:metal_table}, we present the median metallicity of all regions and the elliptical apertures used to obtain these values, allowing us to compare the metallicity between the different regions. With the metallicity from the E region, we can compare it with its stellar mass and see if it follows the MZR as an argument for being a galaxy. The metallicity of the E region is $8.46 \pm 0.16$, and from the 3.4$\mu m$ method using JWST/NIRCam F356W, we obtain its stellar mass of $2.6 \pm 0.4 \times 10^{9} \text{M}_{\odot}$. Using the \citet{Curti2019} calibration of the MZR, this stellar mass corresponds to a gas metallicity of $8.6 \pm 0.07$, consistent with the metallicity obtained with the N2S2 method.}
\textsc{Now, comparing the metallicities of the other regions, including those in the tidal tail of \zwA, which appear to point toward region E (see white arrow in Figure~\ref{fig:mom0_halpha}) and could suggest a physical connection, we find that the metallicity of region E is consistent with the rest within the uncertainties. Therefore, metallicity cannot be used as evidence to support the identification of region E as a separate galaxy in the merger. However, the fact that the E region follows the MZR supports the claim that this is a separate galaxy.}

%%%%%%%%

%%%%%%%%%%%%%%%%%%%%%%%%%%%%%%%%%%%%%%%%%%
%%%% Talk about AGN
%%%%%%%%%%%%%%%%%%%%%%%%%%%%%%%%%%%%%%%%%%
\subsection{Evidence for an AGN in D1}

In this section, we search for the existence of an obscured AGN in D1 based on available multi-wavelength data. It is important to note that some previous studies could not spatially resolve the D1 region and, hence, were contaminated by surrounding extended emissions. \citet{Baan2006} classified \zw as a pure SB using Far Infrared (FIR) (spatial resolution of 15\arcsec) and radio (spatial resolution of 1\arcsec), based on three parameters: the FIR -radio ratio, brightness temperature ($\text{T}_b$) and the radio spectral slope ($\alpha$). Sources like D0, located at $\sim$ 0.5\arcsec of D1,  contributed to the emission, thus contaminating its classification.

\citet{Vardoulaki2015} introduced a method using $\alpha$ maps measured at 1.49 and 8.44 GHz to classify highly obscured sources. AGN-dominated systems show central $\alpha$ values $<$0.5, with surrounding values $>$1.1, indicative of energy loss. \citet{Wu2022} applied this technique to the D region. The $\alpha$ maps, derived from the 3 GHz and 9 GHz VLA bands with a resolution of 1\arcsec$\times$1\arcsec, show values up to 1.1 at the edges, around $\sim$2\arcsec from the D region, and approximately 0.5 in the center, placing D1 at the boundary between AGN-dominated and SB/AGN classifications.

\citet{Ricci2021} concluded that the X-ray spectrum of \zw\ is well-fitted by an SF model, but the contribution from the whole system makes this classification unreliable for D1. In addition, the NuSTAR non-detection leaves the possibility that D1 could host a heavily obscured AGN, likely Compton thick.

JWST observations have provided unprecedented spatial and spectral resolution. In the mid-IR, \citet{Inami2022} classified D1 using the F770W/F560W color, a proxy for the 6.2 $\mu$m PAH EQW. This diagnostic distinguishes SB, characterized by a high EQW due to strong PAH emission, from AGN, which exhibits lower equivalent width values due to hot dust continuum emission and PAH destruction by hard radiation. D1 falls in the middle of this classification; therefore, an SB is as consistent as a highly obscured AGN. \citet{Garcia-Bernete2024} modeled NIRSpec and MIRI/MRS spectra to find that D1 likely contains a buried AGN with a significant contribution from star formation. Consistently, \cite{GB2025} developed a shorter-wavelength IR diagnostic using PAH EW $6.2 \;\mu m / 3.3 \;\mu m$ vs. the continuum ratio $3 \;\mu m / 5 \;\mu m$. The PAH EW decreases in deeply obscured nuclei due to dust absorption, while the continuum ratio rises steeply in buried AGN, creating a mid-IR bump. In this diagnostic, D1 falls in the obscured nucleus region, indicating a highly obscured AGN. Using the same data, \citet{Donnan2024} used a differential extinction model, revealing strong SF activity and a hot, isolated dust component at $T \sim 1000$ K, likely driven by AGN heating. These IR models suggest that D1 contains a combination of both AGN and SF contributions.

Previous observations by \citet{BarcosMunoz2017} used high-resolution VLA 33 GHz data to characterize D1, reporting a size $<$70 pc. These observations provide a crucial probe of the SF activity in the system, revealing a highly compact SB with an SFR surface density consistent with the maximum starburst scenario proposed by \citet{Thomson2005}. It is argued that the most extreme ULIRGs may represent ``Eddington-limited'' star-forming systems, or ``maximum starbursts,'' where SF occurs at the maximum rate allowed by the dominant feedback mechanism, i.e., radiation pressure on dust. When this limit is exceeded, a common inference is that a significant fraction of the luminosity in the system may originate from an AGN. Given its high SFR surface density and compact nature, D1 likely experiences significant radiation pressure effects, which could drive outflows or suppress further star formation. In this work, regardless of the assumed gas fraction in the calculations or whether supernova feedback is considered, D1 consistently appears as a super-Eddington system, suggesting the potential presence of an AGN from this emission excess. 

Here, we present optical evidence further supporting the presence of an AGN, which, combined with ALMA and Chandra data, suggests the existence of an accreting SMBH and vigorous star formation. \textsc{Extinction has a minimal effect on the BPT diagram because it uses emission line ratios that are close in wavelength, reducing the impact of differential extinction. Therefore, even the high optical extinction observed in D1 (Section~\ref{sec:Extinction}) has only a minor influence on the diagnostic results.} The outstanding spatial resolution of the NFM allows us to dissect the ionization of D1, revealing some areas classified as Seyfert. Whereas, in the WFM, D1 is blended with the brighter, less extinct D0, leading to an overall SF classification (yellow point in Figure \ref{fig:Bpt_nii}, \ref{fig:Bpt_sii}, \ref{fig:Bpt_oi}). This is most clearly observed in the [N II] diagram, where the increase of $\log(\text{[O III]/H}\beta)$ is characteristic of a hard photoionization source such as an AGN \citep{Baldwin1981, Kewley2001}. The D1 region, represented by the yellow point in the [S II] and [O I] diagnostic diagrams, is also consistent with an AGN classification when considering the error bars. \textsc{A possible explanation for the ionization in region D1 could be a PNe, as its line ratios are consistent with those observed in such objects according to the original BPT diagram \citep{Baldwin1981}. However, the most luminous PNe reach [O\,\textsc{III}] luminosities of at most 2 $\sim  10^{39} \: \mathrm{erg\,s^{-1}}$ \citep{Ciardullo2002}. As discussed in Section~\ref{sec:Ionization source}, we obtained the spectrum of D1 and, after correcting for optical extinction, derived an [O\,\textsc{III}] luminosity of $\sim 10^{42} \: \mathrm{erg\,s^{-1}}$, far exceeding the expected range for PNe, and hence arguing against this scenario. This value could help classify how strong of an AGN D1 is, based on its [O\,\textsc{III}] luminosity \citep{Kauffmann_sdss_2003}. However, the contribution from star formation could not be reliably quantified, so a definitive classification as weak or strong remains uncertain.}

This result supports various studies in the infrared, which conclude that D1 is a combination of an SB and a heavily obscured AGN. For instance, \citet{Donnan2024}, using a differential extinction model with NIRSpec + MIRI JWST data, conclude that the AGN is affected by higher dust obscuration levels in comparison to the SB, meaning that the optical diagnostics should be dominated by this high SF component, placing D1 near the edge of the Seyfert classification as observed. To test if non-AGN ionization is capable of producing the line ratios observed in D1, we use a pure SB model with \texttt{CLOUDY} \citep{Ferland2013} in combination with shocks emission modeled using \texttt{MAPPINGS III} \citep{Allen2008}. We refer to Appendix \ref{sec:Appendix A} for the details of the two models and the parameters used. The combined ionization sources cannot reproduce the observed line ratios in the D1 region, meaning that an additional ionization source is required, thus supporting the presence of an (obscured) AGN.

%%%%% ALMA
The ALMA Band 3 continuum emission (84–116 GHz) can trace thermal dust emission. However, considering D1's compact size (as shown in Figure \ref{fig:Bpt_nii} from the ALMA Band 3 emission and as measured by \citealp{Inami2022}), the radiation could also originate from synchrotron self-absorption in the nuclear plasma due to its compactness and high density, which are related to an AGN \citep{Kawamuro2022, Ricci2023}. For this highly dense and obscured source, observations at lower frequencies (e.g., Band 1) would be necessary to classify the source using the spectral slope and confirm the presence of an AGN. If we assume that all the 100 GHz emission comes from the AGN, we can compute an upper limit for the column density, $\text{N}_{\text{H}}$, by applying the $\text{F}_{2-10\text{:keV}}/\text{F}_{100\text{\:GHz}}$ relation from \citet{Ricci2023}. Using the X-ray flux from the C+D region \citep{Ricci2021} and the 100 GHz flux from ALMA Band 3 (\(S_{\nu} = 1.4 \times 10^{-3} \, \text{Jy}\)), we obtain a ratio \(< 2.7\), suggesting log N$_{\text{H}} (\text{cm}^{-2}) \; > 24.5$, being evidence of a highly obscured AGN, consistent with the results obtained by \cite{Garcia-Bernete2024} of a log N$_{\text{H}} (\text{cm}^{-2}) \; \sim$ 24.1 - 24.4. However, this value should be considered an upper limit, as dust may significantly contribute at 100 GHz, given that this region has been identified as highly compact with a large amount of dust \citep{Inami2022}.

%%%%% Chandra
In the Chandra X-ray data, D1 has too few counts to generate a statistically meaningful spectral fit. For this reason, we rely on the HR instead. D1 has an HR of 0.1$\pm$0.3, which suggests it is consistent with a hard X-ray source (HR $>$0), a characteristic of the X-ray spectrum of an (obscured) AGN. However, this HR determination is not statistically significant, and it could be due to the dominance of emission from an obscured star-forming region. Moreover, the non-detection by NuSTAR does not rule out the presence of a heavily obscured AGN, likely Compton thick.

Summarizing, the high-resolution [NII], [SII], and [OI] diagnostic diagrams from the NFM, combined with the ALMA and Chandra data, support the presence of an obscured AGN in D1, possibly coexisting with a highly dense SB. The violent interaction between C+D and E has heavily disrupted this region, causing a significant loss of angular momentum in gas and dust, which may fuel the SMBH in D1. In particular, Figure \ref{fig:mom1_halpha} shows a redshifted stream extending north of the D region, likely the source of the inflowing material feeding the nuclear activity in D1.

\subsection{Merger stage}
\begin{figure*}[!htb]
    \centering
    \includegraphics[width=1\linewidth]{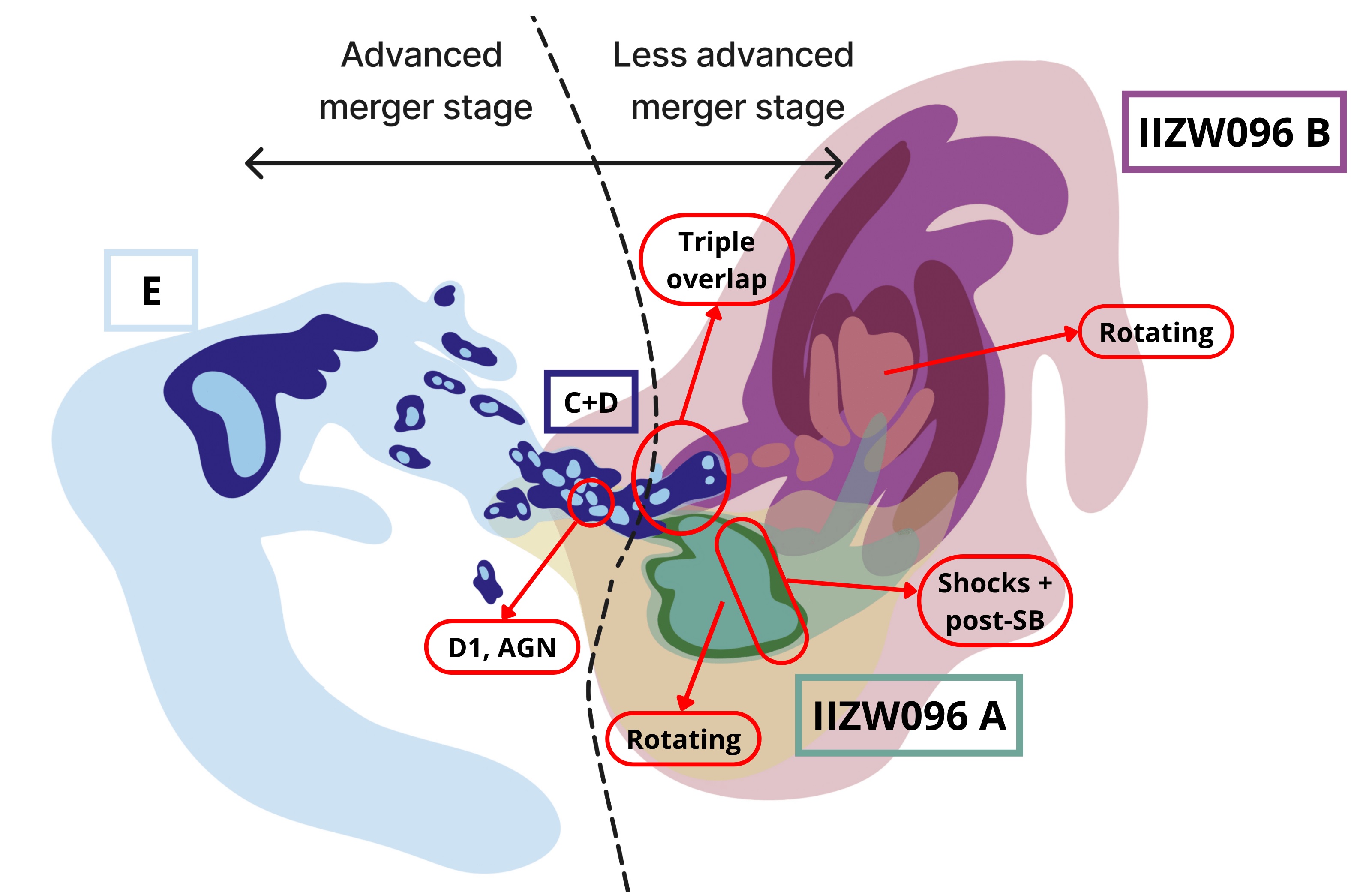}
    \caption{Representation of the merger stage of \zw. \zwA is shown in green, \zwB in magenta, and the C+D and E regions are represented in blue. \zwA exhibits two main interactions: the prominent tidal tail extending toward C+D, characterized by shocks and the post-SB event, and another interaction directed toward \zwB. To the east of this galaxy lies the triple-overlap region, which contains multiple star-forming clumps from the three main systems. To the north is \zwB, with its prominent spiral arms, indicating that it is less perturbed by the merger. To the east are the C+D regions and the obscured AGN in D1, while between C+D and E, filaments connect these regions, remnants of their interaction.
}
    \label{fig:concl}
\end{figure*}

In Figure \ref{fig:concl}, we combine the system's morphology, kinematics, and ionization analyses presented in this work. We propose that the current merger stage of \zw\ involves at least three systems, further divided into two distinct merger stages. 

The west side of the system, which includes \zwA and \zwB, remains in the initial stage of the fusion. Despite being involved in the merger, as evidenced by the numerous SF clumps, tidal tails \citep{Larson2019} and possible outflows from high SF, their morphology and kinematics remain relatively undisturbed, still exhibiting spiral arms and rotation in each disk. With the stellar mass and SFR, we obtain the difference of each component to the main sequence (MS) for $z \sim 0$ \citep{Elbaz2007}. \zwA has a $\Delta \text{MS}=9$ and \zwB a $\Delta \text{MS}=13$, being SF galaxies, occupying  the same region as local (U)LIRGs (e.g., see Figure 12 of \citet{Eser2014})

The merger's east side comprises the E and C+D regions. The stellar mass of the C+D region has not been obtained, but by comparing it to E with \zwA and \zwB, a mass ratio of 8.5$\pm$2 and 17$\pm$4 is derived, respectively. This indicates that the E region side of the merger represents a minor merger (with a mass ratio $<$1:3). For this side of the system, we propose two different scenarios. First, E and C+D were part of the same galaxy, and an interaction with \zwA and/or \zwB tore this galaxy apart, causing significant disruption and separating the SMBH in D1 from the rest of the galaxy, leaving the E region behind. The second scenario is that E and C+D are separate systems. Due to low orbital circularity ($\epsilon$), a direct collapse between them resulted in a straight-line plunge interaction, causing C+D to be highly disrupted \citep{Solanes2018}. In both scenarios, the velocity gradient in E could arise from a rotating disk of the galaxy or from a tidal arc, where the northern part shows receding velocities (redshifted). In contrast, the arc changes direction and becomes approaching (blueshifted).

This side of \zw is more advanced in the merger stage, where the interaction it undergoes triggers a rapid starburst episode in E, evidenced by its prominent H$\beta$ absorption lines and its position below the MS, with a $\Delta \text{MS}=-0.4$. Additionally, the interaction causes significant disruption around the SMBH in D1, where the gas and dust lose their angular momentum, feeding the SMBH and igniting the AGN, a process commonly observed in advanced stages of interaction \citep{Kocevski2015}, where a possible inflow feeding D1 could originate from the redshifted stream in the north direction of D.

The \zw\ system resembles a compact galaxy group (CG). These are associations of 3 to 10 galaxies within projected radii of a few tens of kiloparsecs \citep{Sohn2016}. \cite{Tzanavaris2010} analyzed a sample of local CGs and found that their locations on the MS do not differ much between each group member. This suggests that \zw\ is a unique system where the different galaxies are distributed across the MS. In this system, \zwA and \zwB are classified as SB galaxies, while the E region is identified as inactive. This configuration is reminiscent of the dual merger Mrk 266, where the two nuclei are in different stages of the merger process \citep{Ruby2024}. As the merger evolves, the different components will move closer to nuclear coalescence. The gas and dust will lose angular momentum, further increasing SF and possibly igniting additional AGNs \citep{Foord2020}. This process could reduce the SMBH merger time by a factor of 10 compared to a binary merger \citep{Blaes2002}, generating an even more violent and short-lived phase. Millennium simulations \citep{Ryu2018} found that 42\% of massive galaxies ($10^{11-12} M_{\odot}$) undergo more than one significant merger. This indicates that systems like \zw\ are the predecessors of some of the most massive galaxies in the universe, and observing these rare phases in the local universe is essential for understanding the formation and evolution of galaxies.

\section{Conclusions}
\label{sec:Conclusion}

A detailed analysis of the morphology, kinematics, and ionization of the merging LIRG \zw\ was conducted using VLT/MUSE optical IFU observations. Thanks to the large FoV of the WFM and the high spatial resolution of the NFM, we have been able to resolve and understand better this complex system, its merger stage, and the ionization of its different components. By fitting the main optical emission lines, we obtained flux, velocity, and velocity dispersion maps, as well as optical diagnostic diagrams to characterize the system \textsc{and compare it with HII + shock models, in order to distinguish the contribution of each photoionization source}. We concluded that \zw comprises three or more galaxies. The H${\alpha}$ flux maps identify two central galaxies, \zwA and \zwB, still recognizable as spirals and less affected by gravitational interactions. Multiple clumps have been identified, associated with both spiral arms and collision regions, such as the triple overlap region, which exhibits higher extinction and star formation indicative of compressed molecular gas. Combining the flux and kinematics maps, we separated various double-peaked emission line regions associated with interacting zones to model their kinematics, obtaining a PA = 99$^\circ$ $\pm$ 24$^\circ$ and $i$ = 40$^\circ$ $\pm$ 21$^\circ$ for \zwA, and a PA = 170$^\circ$ $\pm$ 5$^\circ$ and $i$ = 65$^\circ$ $\pm$ 15$^\circ$ for \zwB. The residuals reveal a tidal tail of \zwA in the direction of \zwB and a region of high density of stars not following the rotation pattern of the galaxy. In addition, to estimate the magnitude of the merger between these two galaxies, the JWST NIRCam F356W filter was used as a proxy for stellar mass, yielding stellar masses of \(2.2 \pm 0.4 \times 10^{10} \: M_{\odot}\) for \zwA and \(4.5 \pm 0.8 \times 10^{10} \: M_{\odot}\) for \zwB, implying that these two galaxies are undergoing a major merger. Furthermore, by separating the emission line components, we were able to analyze the kinematics of a broad component at the center of \zwA. Due to its velocity (-60 to -90 km$/$s) and velocity dispersion ($>$180 km$/$s), there is a possibility that this is a weak outflow from a region of high SF, as suggested by its composite ionization in the [NII] diagram.\textsc{ We thus labeled it as "potential outflow," or PO.}

The east side of the system is highly disrupted and strongly affected by the interaction. The C+D region contains multiple bright clumps with high extinction, reaching A$_{H\alpha}\sim$5.3 in D1, consistent with its high L$_{IR}$. We also identified the E region, a faint system connected to the C+D region, visible in the flux map and the color image. The kinematics in the C+D region are very complex, revealing multiple double-peaked lines due to structures with different velocities in the same FoV, likely residuals from the interaction. The E region shares a velocity gradient across different lines with the same PA, suggesting a possible rotating disk.

From the resolved optical line diagrams, the dominant classification of the system falls under SF, except for the colliding regions. The tidal tails of the \zwA show increased ionization, \textsc{evidenced by the match with the HII + shock models with higher shock contributions, also supported by the high soft X-ray emission detected by Chandra. In addition, these regions show a prominent H$\beta$ absorption line, where hot young stars and possibly more evolved and hot stars could also be providing the photons to increase the ionization. This means that these regions show an increase in ionization from shocks plus a post-SB event, meaning that this galaxy is more affected by gravitational forces in comparison to \zwB, where there is no evidence of a higher shock contribution. In addition, the compact sources of this galaxy are also best matched by HII regions with high $\log(Q(H))$ between 7.25 and 8, whereas \zwB is better matched with HII region models with lower $\log(Q(H))$, between 6.75 and 7.25. The C+D region shows a similar behavior, where the compact sources are better fitted by HII models with higher $\log(Q(H))$, between 7 and 8, and also exhibit a higher shock contribution. This, combined with its complex morphology and kinematics, suggests a past interaction. Moving toward the E region, we can trace the remnants of this interaction through the increase in ionization observed in the [OI] diagram, attributed to the lower ionization potential of this line.}

By combining the diagnostic diagrams, ALMA, and Chandra observations, we argue that D1 contains a highly obscured AGN in combination with a compact starburst. This is supported by the optical line classification of D1 as an AGN, \textsc{and by its high L$_{[OIII]}$, of $\sim 10^{42}$ erg$/$s}. Given its compactness and density, the ALMA Band 3 continuum indicates that this compact source could be attributed either to thermal dust emission or to synchrotron self-absorption associated with an obscured AGN. Using Chandra data, we computed the Hardness ratio (HR), resulting in a positive value (-0.56 $\pm$ 0.08) consistent with a hard source and supporting the presence of an obscured AGN. These three lines of evidence make a strong case for an (obscured) AGN in D1.

Combining the morphological, kinematic, and ionization information, we propose the merger stage depicted in Figure \ref{fig:concl}, highlighting the two evolutionary stages. In a relatively early merger stage, the west side shows remnants from the interaction, such as tidal tails, SF clumps, and shocks in the interacting regions, while conserving its morphology and kinematics, being classified as SB galaxies. On the other side, the east side is at a more advanced stage. It consists of two systems, C+D and the E regions, where we proposed two scenarios. First, E and C+D are part of the same galaxy, disrupted by an interaction with \zw and/or \zwB. The other scenario is that E and C+D are isolated galaxies, and from a direct interaction, C+D gets highly disrupted, fueling the SMBH in D1. This interaction rapidly quenched the E region in both scenarios, with its SF phase occurring on a short timescale. As the system evolves, a new reservoir of gas and dust may further ignite activity, increasing SF and possibly triggering additional AGN(s). 

\textsc{The richness and complexity of this nearby system is a clear example of the diverse and somewhat chaotic nature of these multiple galaxy interactions, which could be essential in understanding the formation and evolution of the most massive galaxies in the Universe. In this work, we studied in detail the highest resolution optical IFU data cubes available to date of a relatively nearby system, demonstrating how much more challenging it would be to perform a similar analysis at higher redshifts, where such multiple encounters are expected to be more frequent.}

% Adds a space of 1 cm
\section*{Acknowledgements}
This work is based on observations collected at the European Southern Observatory under ESO program 097.B-0427 and 109.23AX.00.
This research used data obtained from the Chandra Data Archive provided by the Chandra X-ray Center (CXC).
This paper makes use of the following ALMA data: ADS/JAO.ALMA\#2017.1.01235.S. ALMA is a partnership of ESO (representing its member states), NSF (USA), and NINS (Japan), together with NRC (Canada), NSTC and ASIAA (Taiwan), and KASI (Republic of Korea), in cooperation with the Republic of Chile. The Joint ALMA Observatory is operated by ESO, AUI/NRAO and NAOJ. This research is based on observations made with the NASA/ESA Hubble Space Telescope obtained from the Space Telescope Science
Institute, which is operated by the Association of Universities for Research in Astronomy, Inc., under NASA contract NAS 5–26555.
These observations are associated with programs 10592 and 17285. This work is based [in part] on observations made with the NASA/ESA/CSA James Webb Space Telescope. The data were obtained from the Mikulski Archive for Space Telescopes at the Space Telescope Science Institute, which is operated by the Association of Universities for Research in Astronomy, Inc., under NASA contract NAS 5-03127 for JWST. These observations are associated with program 1328. 

\textsc{We thank the anonymous referee for the very insightful and detailed comments and suggestions that significantly improved our original manuscript.} \textsc{We thank Drs. Jeff Rich and Mike Dopita for sharing valuable data from their shock ionization models.} We acknowledge support from ANID programs BASAL FB210003 (CATA), FONDECYT Regular 1200495, 1241005, and 1250821, FONDECYT postdoctoral fellowship 3220751 (CF), Millennium Science Initiative AIM23-0001 (FEB), and FONDECYT Postdoctorado 3200802 (GV). AMM acknowledges support from the NASA Astrophysics Data Analysis Program (ADAP) grant number 80NSSC23K0750. V.U acknowledges funding support from NSF Astronomy and Astrophysics Grant (AAG) No. AST-2408820, NASA Astrophysics Data Analysis Program (ADAP) grant No. 80NSSC23K0750, and STScI grant Nos. HST-AR-17063.005-A, HST-GO-17285.001-A, and JWST-GO-01717.001-A. F.M-S. acknowledges support from NASA through ADAP award 80NSSC19K1096 and through award 80NSSC23K1529. CR acknowledges support from Fondecyt Regular grant 1230345 and the China-Chile joint research fund. CRA acknowledges support from the project ``Tracking active galactic nuclei feedback from parsec to kiloparsec scales'', with reference PID2022-141105NB-I00. D.T-A. acknowledges support by DLR grant FKZ 50 OR 2203. GD acknowledges support by UKRI-STFC grants: ST/T003081/1 and ST/X001857/.  PBT acknowledges partial funding by Fondecyt-ANID 1240465/2024. GV and SZ acknowledge support by European Union’s HE ERC Starting Grant No. 101040227 - WINGS. SZ acknowledges support from the Spanish grant from the present Ministry of Science and Innovation through the research grant PID2019-107408GB-C42. SA gratefully acknowledges support from the ERC AdG project 789410 and from the Knut and Alice Wallenberg project ”The Origin and Fate of Dust in the Universe”. GO gratefully acknowledges support from the Knut and Alice Wallenberg project ”The Origin and Fate of Dust in the Universe”

% Appendix Title Spanning Both Columns
%\bibliography{introduccion2.bib}{}

\newpage
%\twocolumn
\begin{appendix}
\section{Appendix: Emission Line Ratios}
\label{sec:Appendix Line Ratio}

\begin{figure}[H]
    \centering
    \includegraphics[width=1\linewidth]{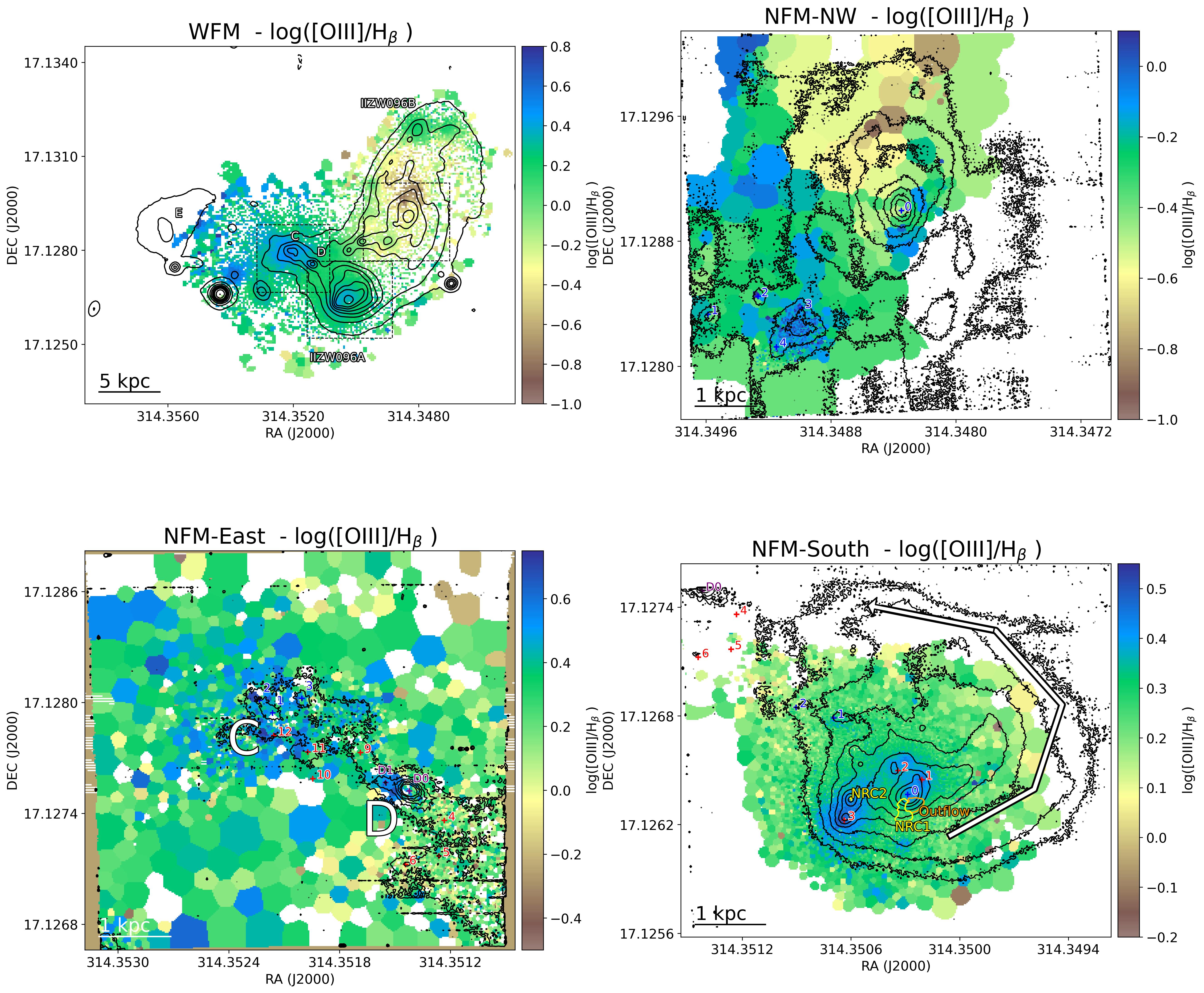}
    \caption{ Emission line flux ratio of $\log([\text{O III}]\,\lambda5007/\text{H}\beta)$. The contours, regions, and sources are defined in Figure \ref{fig:mom0_halpha}, in addition to the PO, NRC1 and NRC2 of \zwA }
    \label{fig:lr_oiii}
\end{figure}

\begin{figure}[H]
    \centering
    \includegraphics[width=1\linewidth]{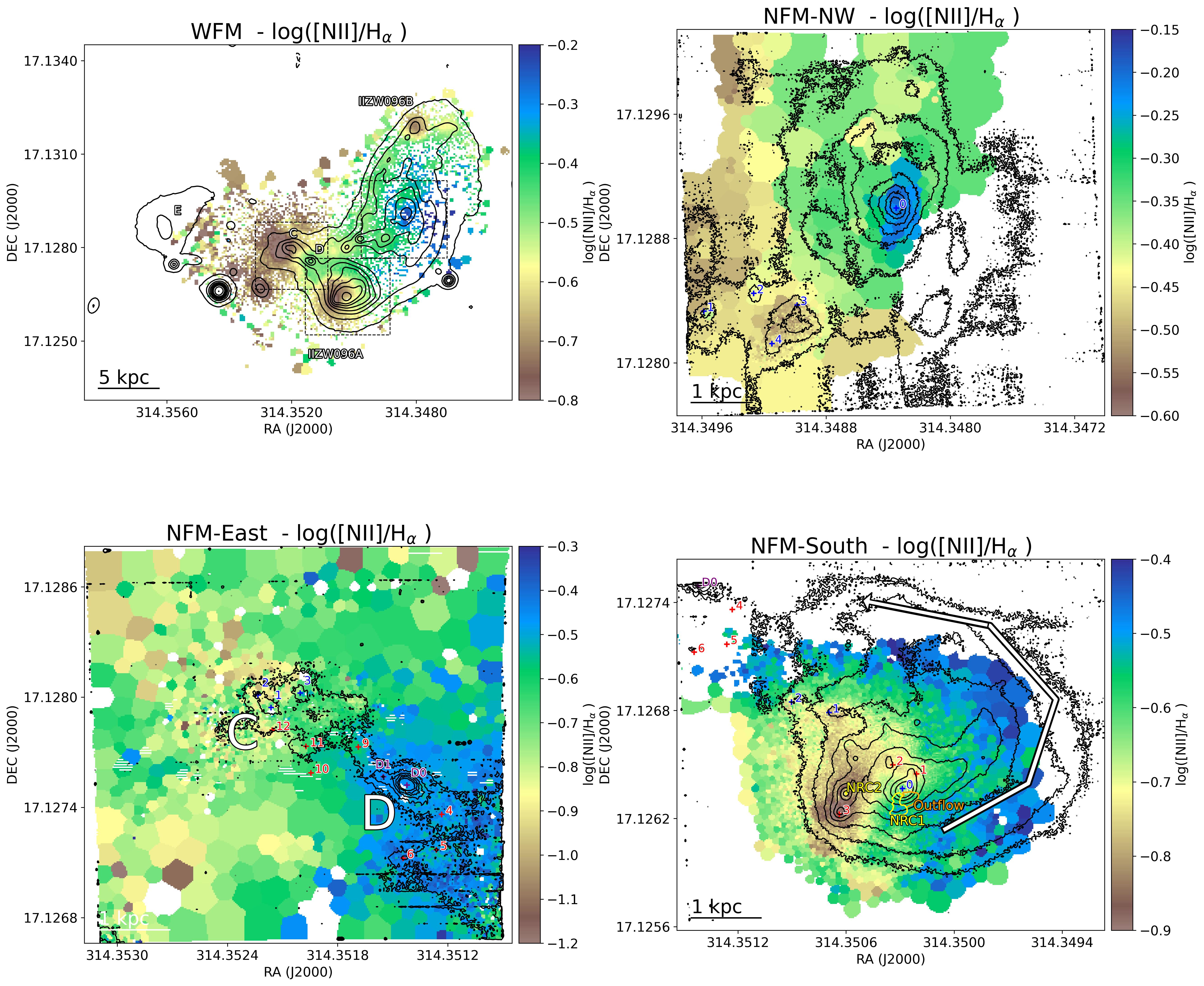}
    \caption{ Emission line flux ratio of $\log([\text{N II}]\,\lambda6583/\text{H}\alpha)$. The contours, regions, and sources are defined in Figure \ref{fig:mom0_halpha}, in addition to the PO, NRC1 and NRC2 of \zwA }
    \label{fig:lr_nii}
\end{figure}

\begin{figure}[H]
    \centering
    \includegraphics[width=1\linewidth]{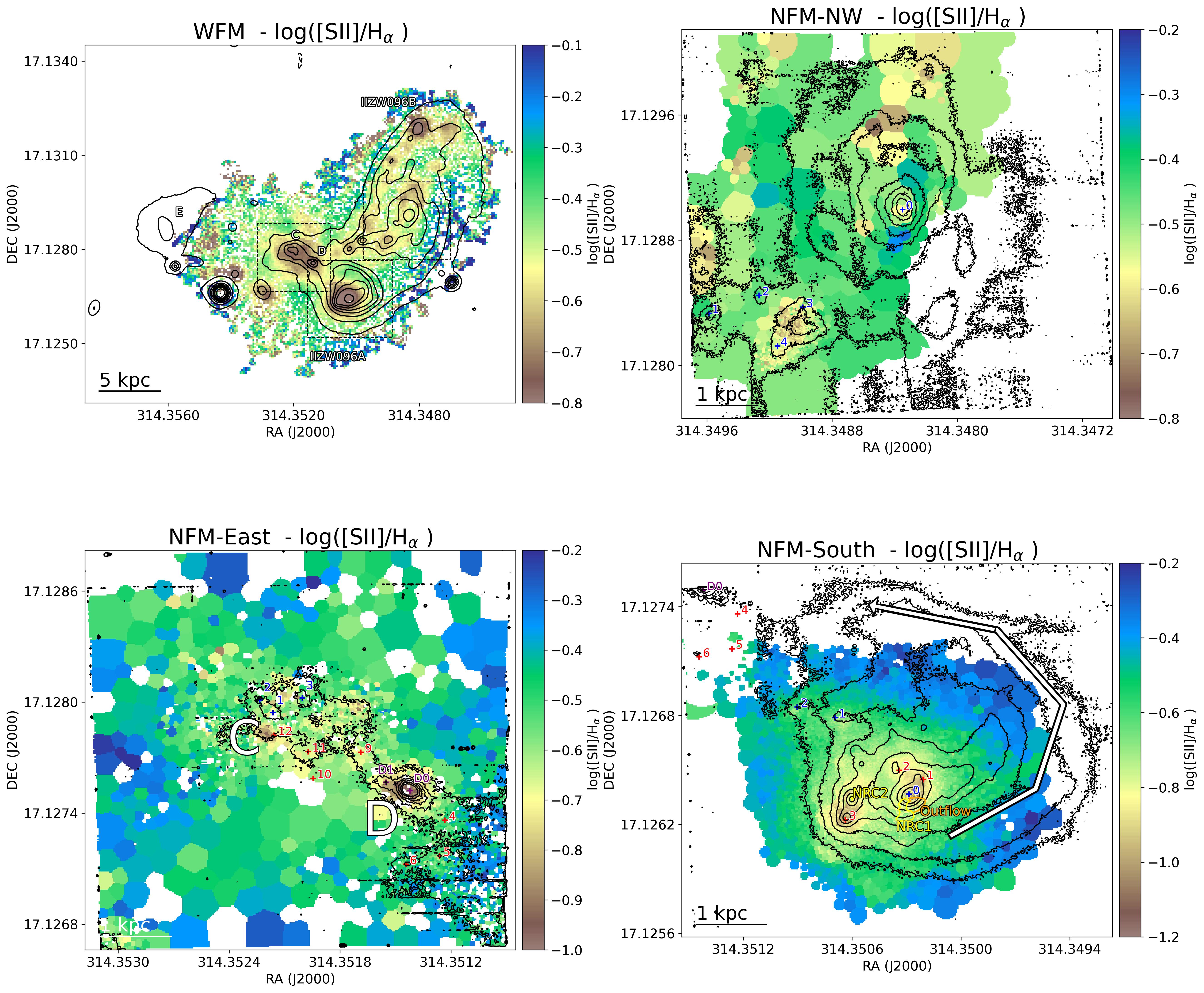}
    \caption{ Emission line flux ratio of $\log([\text{S II}]\,\lambda(6717+6731)/\text{H}\alpha)$. The contours, regions, and sources are defined in Figure \ref{fig:mom0_halpha}, in addition to the PO, NRC1 and NRC2 of \zwA }
    \label{fig:lr_sii}
\end{figure}

\begin{figure}[H]
    \centering
    \includegraphics[width=1\linewidth]{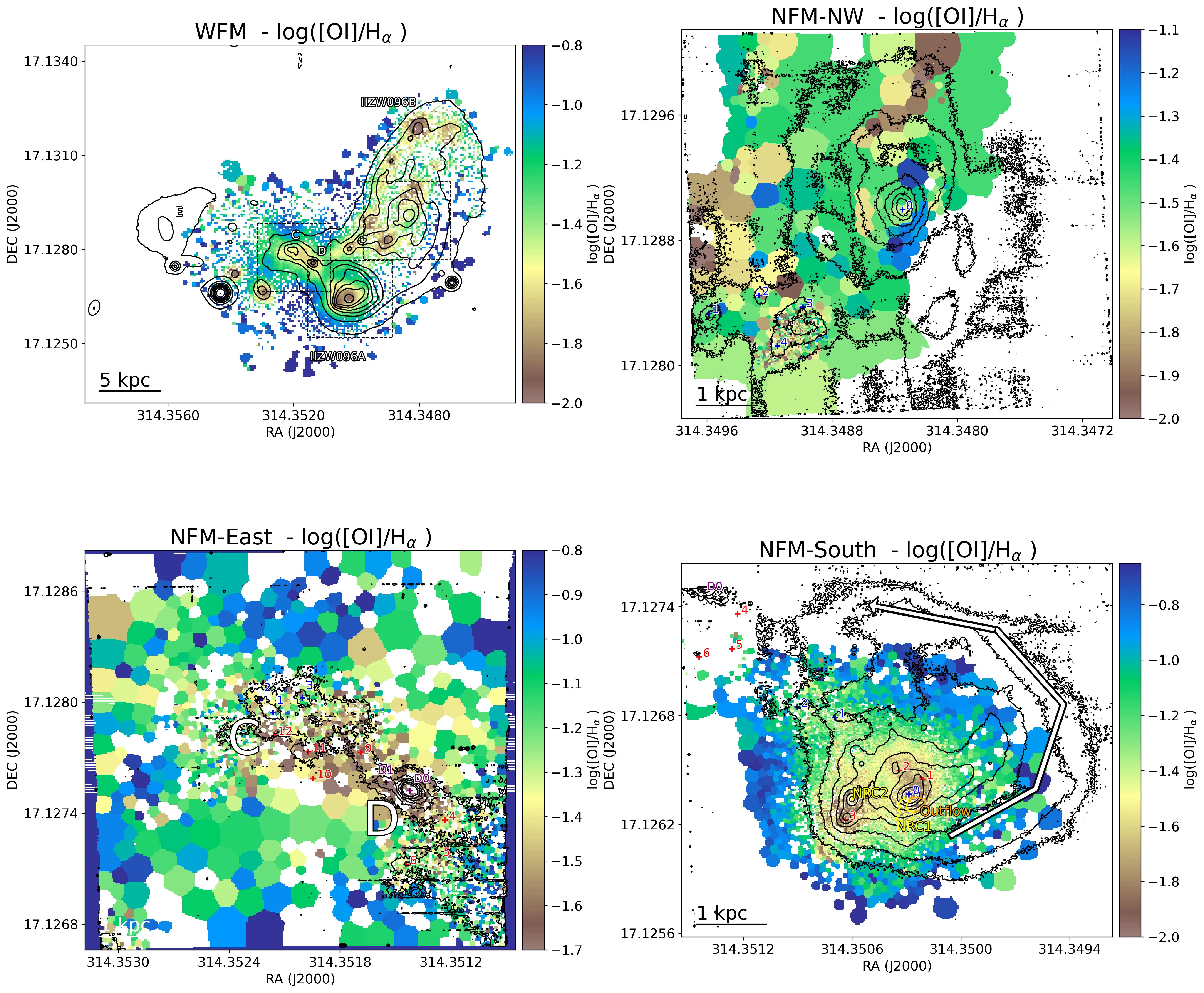}
    \caption{ Emission line flux ratio of $\log([\text{O I}]\,\lambda6300/\text{H}\alpha)$. The contours, regions, and sources are defined in Figure \ref{fig:mom0_halpha}, in addition to the PO, NRC1 and NRC2 of \zwA }
    \label{fig:lr_oi}
\end{figure}

\section{Appendix: Blue bump}
\label{sec:Appendix blue bump}
\begin{figure}[H]
    \centering
    \includegraphics[width=1\linewidth]{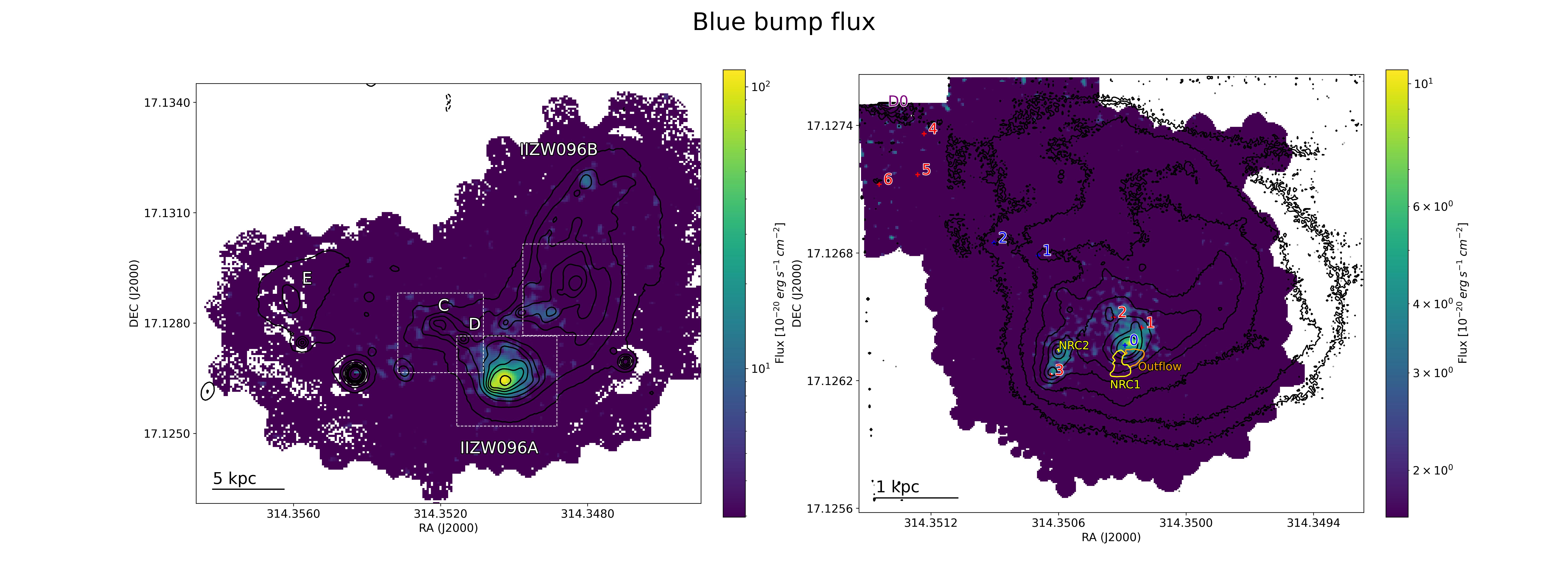}
    \caption{ Map of the blue bump excess characteristic of WR stars, measured in the 4650–4707 \AA spectral range where this feature is identified. Only the NFM-South pointing is shown, as it is the only one exhibiting this signature. The WFM map is included for reference to illustrate that \zwA is the only galaxy with prominent emission of this feature. The contours, regions, and sources are defined in Figure \ref{fig:mom0_halpha} }
    \label{fig:wr}
\end{figure}

In Figure \ref{fig:wr} we present a map of the "blue bump", a characteristic spectral feature of Wolf-Rayet (WR) stars. This map was constructed by collapsing the datacube within the spectral range 4650--4707 \AA, which contains the blue bump emission, and then subtracting the continuum contribution from an adjacent spectral window of equal width (4707--4764 \AA). This subtraction technique isolates the WR feature by removing the underlying stellar continuum and ensures that we are specifically tracing the WR population rather than continuum-dominated regions. We show the results for both the WFM and the NFM-South, with the latter pointing exhibiting the most prominent blue bump features. 

 \section{Appendix: Emission Line Ratio Modeling of D1}
\label{sec:Appendix A}

\begin{figure}[H]
    \centering
    \includegraphics[width=1\linewidth]{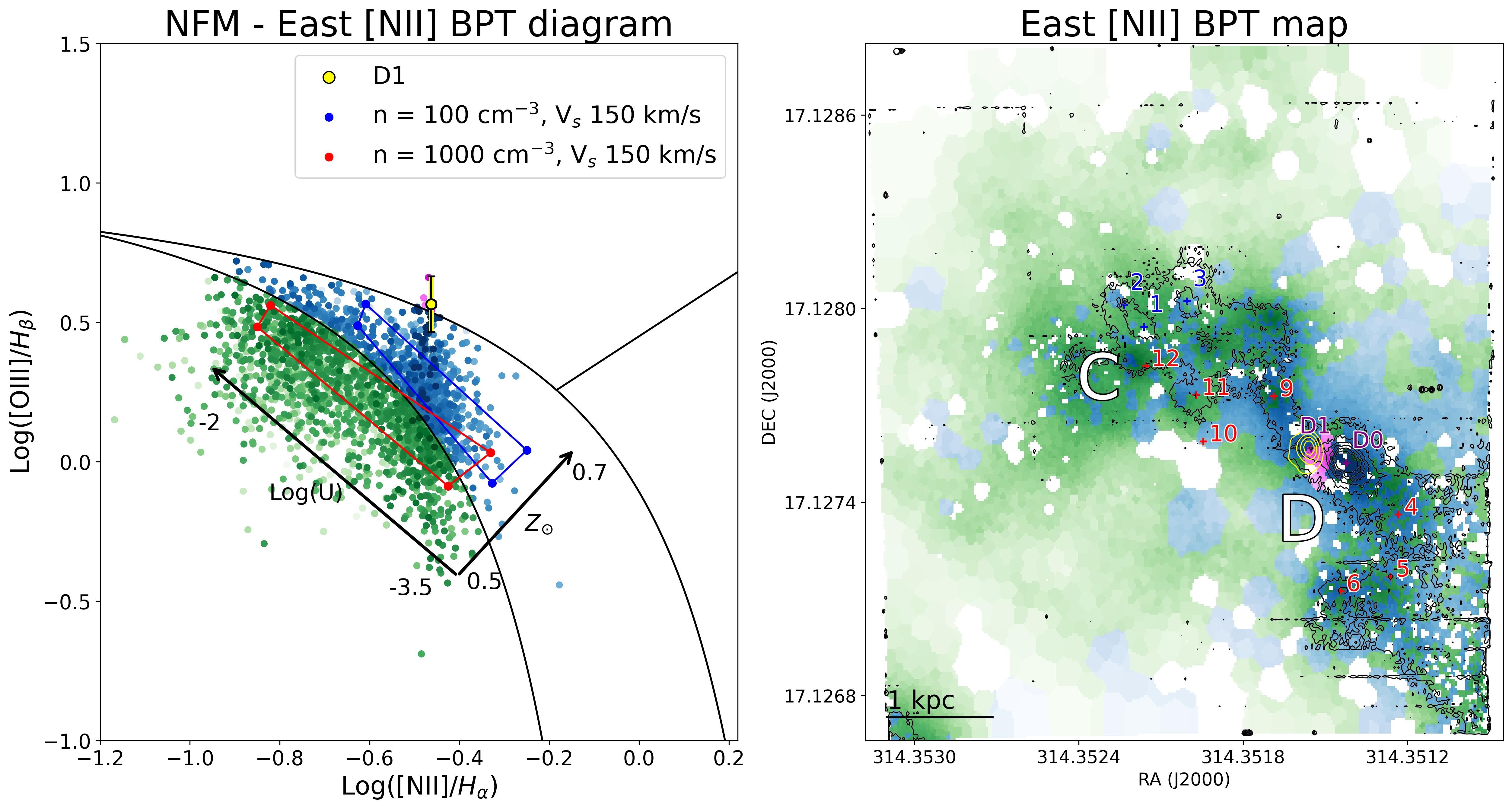}
    \caption{[N II] diagnostic diagram as presented in Figure \ref{fig:Bpt_nii}, with the emission line ratios produced by \texttt{CLOUDY} + \texttt{MAPPINGS III}. A 5 Myr starburst (SB) is synthesized using PopStar (Mollá et al. 2009), assuming a gas metallicity between 0.5–0.7 Z$_\odot$. Solar abundances are assumed for the shocks, with a velocity of 150 km$/$s, $B$=10$\ \mu\text{G}$, and a pre-shock density ranging from 100 to 1000 cm$^{-3}$. The results are represented in red and blue, with variations in the ionization parameter log(U) and the metallicity of the gas in the SB. The observed line ratios in the D1 region are shown in yellow.
}
    \label{fig:cloudy_nii}
\end{figure}

We use a pure SB model with shocks to test whether this ionization mechanism can reproduce the optical line ratios observed with MUSE-NFM. The choice of excluding an AGN is due to the lack of prior studies providing a range of values for key AGN parameters, such as the ionization parameter, N$_{H}$, and hydrogen density. In contrast, for the SB case, previous studies \citep{Inami2010} have characterized the stellar population D1 age, a key parameter for modeling the line ratios. Without similar constraints for an AGN, its inclusion would introduce significant degeneracies. Thus, we first assess whether an SB+shock model alone can explain the observations.

For the SB, we use \texttt{CLOUDY} \citep{Ferland2013}, a radiative transfer simulation code capable of reproducing line ratios under specific conditions in a non-equilibrium gas. The ionization is driven by a young SB synthesized with PopStar \citep{Molla2009}, using a Kroupa IMF with lower and upper mass limits of 0.15–100 M$_{\odot}$. We select an age of 5 Myr, based on estimates from the CO index and $Br\ \gamma$ equivalent width \citep{Inami2010}, a gas metallicity between 0.5–0.8 Z$_{\odot}$ derived from the N2 diagnostic \citep{Marino2013} using the NFM observations, and a radius of 134 pc \citep{Wu2022}. With \texttt{CLOUDY}, we generate a grid of models varying the metallicity and the ionization parameter $U$, a dimensionless quantity that represents the ratio of the ionizing photon flux to the gas density, indicating the degree of ionization in a gas cloud.

For the shocks, we use the \texttt{MAPPINGS III} models from \citet{Allen2008}, adopting a velocity ($v_s$) of 150 km$/$s. This choice is based on the results of \citet{Pereira-Santaella2024}, who used MgIV to trace shocks in D1 and concluded that velocities of 150 km$/$s best agree with the observed line ratios. We consider solar abundances with a pre-shock density $n_e$ ranging from 100–1000 cm$^{-3}$. \citet{Wu2022} estimates the density of D1 as $4 \times 10^4 \; \text{cm}^{-3}$, meaning that the values closest to the observations should be from $n$=1000 $\text{cm}^{-3}$. Additionally, we use the maximum magnetic field strength ($B$) available in the models, 10 $\mu G$, to test the most extreme conditions capable of producing higher ionization.

The two models provide line ratios normalized to H$\beta$. To combine the contributions of the SB and shocks, we require the luminosity of one of the lines. For the SB, we use an SFR=60 M$_{\odot}$/yr, as derived from JWST/MIRI observations \citep{Inami2022} and converted it to L$_{H\alpha}$ using the relation from \citet{Kennicutt1998}. From this, we calculate the remaining luminosities of the [N II] diagnostic diagram lines from the line ratios provided by the models. For the shocks, \citet{Rich2015} analyzed local U/LIRGs at different merger stages and found that shocks may contribute up to 50\% of the H$\alpha$ luminosity in late-stage mergers in the absence of an AGN. Here, we assume half of the H$\alpha$ luminosity derived for the SB as the contribution from shocks. However, this should be considered an upper limit, as an AGN can further contribute to this luminosity.

In Figure \ref{fig:cloudy_nii}, we present the non-AGN models produced by combining SB and shocks and compare them with the observed D1 points. The two different colors, red and blue, represent the same SB model varying the ionization parameter $\log(U)$ and metallicity, but with different pre-shock densities (100 and 1000 cm$^{-3}$ ). Both models adopt the same $v_s$ and maximum $B$ provided by \texttt{MAPPINGS III} but differ in pre-shock densities. The ionization from non-AGN sources is expected to lie between the two grids at different densities within the SF and Composite regions of the [N II] diagnostic diagram. However, to reproduce the D1 point, an additional ionization source is required, supporting the presence of an AGN.
  
\end{appendix}

\end{document}